\documentclass[a4paper,11pt]{article}
\usepackage{jheppub}

\usepackage{amsmath}
\usepackage{amsfonts}
\usepackage{amssymb}
\usepackage{graphicx}
\usepackage{wrapfig}
\usepackage{braket}
\usepackage{hyperref}
\usepackage{amsthm,verbatim,cancel}
\usepackage{color}
\usepackage{caption}
\usepackage{subcaption}
\newcommand{\ra}{\rightarrow}
\newcommand{\dg}{\dagger}
\newcommand{\bea}{\begin{eqnarray}}
\newcommand{\eea}{\end{eqnarray}}
\newcommand{\nn}{\nonumber\\}

\DeclareMathOperator{\tr}{tr}

\DeclareMathOperator{\U}{\mathrm{U}}
\DeclareMathOperator{\SU}{\mathrm{SU}}
\DeclareMathOperator{\SO}{\mathrm{SO}}
\DeclareMathOperator{\Sp}{\mathrm{Sp}}

\DeclareMathOperator{\C}{\mathbb{C}}

\newcommand{\ketbra}[2]{\ket{#1}\!\!\bra{#2}}

\begin{document}

\title{Entanglement branes in a two-dimensional string theory}

\author[a]{William Donnelly}
\emailAdd{donnelly@physics.ucsb.edu}
\affiliation[a]{Department of Physics, University of California, Santa Barbara, CA 93106}

\author[b]{Gabriel Wong}
\emailAdd{gw2gt@virginia.edu}
\affiliation[b]{Department of Physics, University of Virginia, Charlottesville, VA 22901}

\abstract{

What is the meaning of entanglement in a theory of extended objects such as strings?
To address this question we consider the spatial entanglement between two intervals in the Gross-Taylor model, the string theory dual to two-dimensional Yang-Mills theory at large $N$.  The string diagrams that contribute to the entanglement entropy describe open strings with endpoints anchored to the entangling surface, as first argued by Susskind.   We develop a canonical theory of these open strings, and describe how closed strings are divided into open strings at the level of the Hilbert space.  We derive the Modular hamiltonian for the Hartle-Hawking state and show that the corresponding reduced density matrix describes a thermal ensemble of open strings ending on an object at the entangling surface that we call an E-brane.}

\maketitle

\newpage

\section{Introduction}

Perhaps the most profound insight about gravity is its holographic nature.
The Bekenstein-Hawking entropy formula
\begin{equation}
S_\text{BH} = \frac{A}{4G}, \label{SBH}
\end{equation}
(in units with $c = \hbar = 1$) is an indication that the relevant degrees of freedom for describing the interior of a black hole --- and possibly any region of space --- reside on its boundary at Planckian density.

The area scaling of the entropy could be naturally explained if the relevant entropy is the entanglement entropy of quantum fields in their vacuum state \cite{Sorkin1983,Bombelli1986,Srednicki1993}.
Given a region of space $V$ and its complement $\bar V$, one introduces a tensor product decomposition of the Hilbert space $\mathcal{H} = \mathcal{H}_V \otimes \mathcal{H}_{\bar V}$, and defines the reduced density operator $\rho_V = \tr_{\mathcal{H}_{\bar V}} \ketbra{\psi}{\psi}$.
The entanglement entropy of the region $V$ in state $\ket{\psi}$ is defined as the von Neumann entropy of the reduced density operator,
\begin{equation}
S = - \tr \rho_V \log \rho_V.
\end{equation}
In continuum field theory this entropy is divergent: it typically follows an area law, but with the ultraviolet cutoff playing the role of the Planck length.
Thus in order to obtain a finite entropy consistent with \eqref{SBH}, we have to incorporate gravity into the entanglement calculation, necessitating a theory of quantum gravity.
Finiteness of the entropy is crucial to the resolution of the information paradox, as otherwise a black hole could in principle store an infinite amount of information.

\begin{figure}
\centering 
\includegraphics[scale=1]{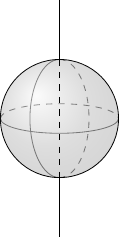}
\qquad \qquad
\includegraphics[scale=1]{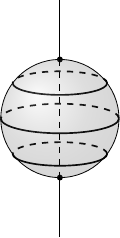}
\qquad \qquad
\includegraphics[scale=1]{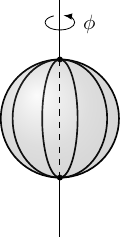}
\caption{
In string theory, $\log Z(\beta)$ is a sum of connected string diagrams embedded in a spacetime with a conical defect of angle $\beta$ at the entangling surface.
Only closed string diagrams that intersect the entangling surface, such as the sphere diagram on the left, contribute to the entanglement entropy.
Sliced transverse to the entangling surface, the middle diagram describes a closed string emitted from a point on the entangling surface and then reabsorbed.
Sliced in angular time $\phi$, the sphere diagram is a one loop open string diagram with the endpoints fixed on the entangling surface as depicted on the right.
}
 \label{figure:SU}
\end{figure}

A leading candidate for such a theory of quantum gravity is string theory, and entanglement in string theory has been studied using the replica trick \cite{Susskind:1993ws,Susskind:1994sm,Dabholkar:1994ai,Dabholkar:1994gg}.
To calculate the entropy via the replica trick one considers the Euclidean path integral $Z(\beta)$ on a spacetime where the angular coordinate $\phi$ is identified with period $\beta$.
When $\beta \neq 2 \pi$, this spacetime is singular, containing a planar conical defect of angle $\beta$.
The replica trick then gives the entropy as
\begin{equation}\label{rep}
S = (1 - \beta \partial_\beta) \log Z |_{\beta = 2 \pi}.
\end{equation}
Ref.~\cite{Susskind:1994sm} showed how the entropy calculation could be organized into contributions from different closed string worldsheets.
It was argued that the genus-zero worldsheets that intersect the conical singularity, one of which is shown in figure \ref{figure:SU}, give the classical Bekenstein-Hawking term, and the torus diagrams give its one-loop correction.
These different diagrams can be foliated by an angular coordinate going around the entangling surface, and describe propagation of both closed and open strings, where the open string endpoints are anchored to the entangling surface.

While the replica trick calculation is very efficient, it is somewhat opaque in that it does not provide a canonical description of the entropy.
This raises an important question: which states are being counted in this calculation?
The replica calculation is further complicated by the fact that it relies on the path integral in the background of a conical defect, for which one would have to define string theory off-shell.

A closely related issue arises in field theory, and it concerns the way that certain fields couple to a conical singularity in spacetime.
For fields with nontrivial coupling to curvature the path integral on a cone contains in ``contact terms'', which are explicit interactions between the curvature and the dynamical fields.
These contact terms arise for scalar fields with nonminimal coupling \cite{Larsen1995,Solodukhin1995} and for gauge fields \cite{Kabat:1995eq}.
String theory contains fields of all spins, and hence an infinite tower of such contact terms \cite{He:2014gva}.
In theories with massless fields of spin $3/2$ or spin $2$ the situation is even worse: the theory cannot be consistently defined on manifolds such as the conical defect that are not solutions of the Einstein equation \cite{Fursaev1996}.

The presence of contact terms obscures the interpretation of the entropy as calculated via the replica trick \eqref{rep}.
Does the replica formula still calculate the von Neumann entropy of a reduced density matrix, or is there a distinct non-entanglement contribution from contact interactions?
Recently it has been understood that the contact terms in abelian gauge theory \emph{do} correspond to a counting of states, but the entanglement entropy has to be suitably interpreted in a gauge theory.
Since the physical Hilbert space does not factorize, one must consider an extended Hilbert space containing degrees of freedom associated with the boundary surface \cite{Donnelly2011}.
In the case of electrodynamics the entropy of these \emph{edge modes} have been shown to give rise to the contact terms in the entropy \cite{Donnelly:2014fua,Donnelly:2015hxa}.
The inclusion of the edge modes resolves the longstanding puzzle of the interpretation of the contact term, and are necessary to restore agreement between the logarithmic divergence and the central charge.\footnote{The apparent negative sign of the contact term found in ref.~\cite{Kabat:1995eq} and emphasized in ref.~\cite{Donnelly2012} is a red herring; the leading divergence depends on the regularization scheme and may have either sign \cite{Donnelly:2015hxa}.}

The decomposition of the Hilbert space in terms of edge modes is best understood for Yang-Mills theory in two spacetime dimensions \cite{Donnelly2014a}.
Two-dimensional Yang-Mills theory has no local degrees of freedom and so one avoids all issues related to ultraviolet divergences.
The theory is almost topological, and can be exactly solved by topological quantum field theory methods.
The partition function of a closed oriented manifold $M$ depends only on its topology (which is encoded in the Euler characteristic $\chi = 2 - 2 G$, where $G$ is the genus) and its total area $A$.
The partition function can be written exactly as \cite{Cordes:1994fc}:
\begin{equation}\label{ZYM}
Z = \sum_{R} (\dim R)^\chi e^{-\frac{\lambda A}{2N} C_2(R)},
\end{equation}
where $\lambda = g_{\text{YM}}^2 N$ is the (dimensionful) 't Hooft coupling.
Here $R$ runs over all irreducible representations of the gauge group, which we will take to be $\U(N)$, with $\dim(R)$ the dimension of each representation and $C_2(R)$ the quadratic Casimir.

Entanglement entropy for general regions can be obtained from applying the replica formula \eqref{rep} to the partition function \eqref{ZYM} \cite{Donnelly:2014gva,Gromov:2014kia}.
To make sense of this formula and to see which states the entanglement entropy counts, we have to understand how to decompose the Hilbert space into regions of space.
In 1+1 dimensions, Hilbert spaces are associated to 1-dimensional manifolds, which may  be circles or intervals.
On a circle, the Hilbert space is the space of square-integrable class functions on the group, i.e. those functions $\psi:G \to \C$ satisfying $\psi(U) = \psi(g^{-1} U g)$.
On an interval the Hilbert space is the space of square-integrable functions on the gauge group, with no restriction to class functions.
This space carries two unitary actions of $G$, corresponding to left and right group multiplication, and which act at the left and right endpoints of the interval, respectively.
Two intervals can be glued together at a common endpoint using the entangling product \cite{Donnelly:2016auv}; we take the ordinary tensor product of Hilbert spaces and quotient by the simultaneous action of right multiplication on the leftmost interval and left multiplication on the rightmost interval.
This allows us to combine two intervals into the Hilbert space of a larger interval, or to glue both endpoints of an interval together, giving the Hilbert space of a circle.
Using this decomposition, we can embed the state of any region into the tensor product of the subregions and the resulting entanglement entropy reproduces the result of the replica trick calculation \cite{Donnelly:2014gva}.

How do we use this insight from Yang-Mills theory to understand entanglement entropy in string theory? 
The key observation we will exploit, due to Gross and Taylor is that in the large-$N$ limit, Yang-Mills theory \emph{is} a string theory \cite{Gross:1992tu,Gross:1993hu}.
The partition function \eqref{ZYM} can be expressed as a sum over maps from a two-dimensional worldsheet into the spacetime manifold $M$, weighted by the Nambu-Goto action.
The worldsheet can have certain prescribed singularities that represent interactions of the strings.\footnote{One can give a string theory interpretation of two-dimensional Yang-Mills theory for gauge groups $\U(N)$, $\SU(N)$ and even $\SO(N)$ and $\Sp(N)$ \cite{Ramgoolam:1993hh}.
These theories differ in the orientability of the worldsheet, and in the types of singularities that can appear.
Here we will focus on a subsector of the $\U(N)$ theory that leads to the simplest string description, but we expect the broad features to generalize.}
There are additional singularities that must appear on manifolds with $\chi \neq 0$ such as the sphere, which Gross and Taylor called $\Omega$-points.
The $\Omega$-points are unlike the other singularities: they are not integrated over, and their total number is fixed by the Euler characteristic of the manifold.
While their existence is demanded from the evaluation of the Yang-Mills partition function, the reason for the $\Omega$-point singularities from the perspective of the string theory path integral is somewhat obscure.

In order to import our understanding of entanglement in Yang-Mills theory into the Gross-Taylor description, we have to describe the theory canonically in terms of a Hamiltonian operator acting on a Hilbert space.
Baez and Taylor showed how to describe the Hilbert space of a circle and its time evolution along a cylinder in the Hamiltonian language as a string field theory \cite{Baez:1994gk}.
We review this picture in section \ref{section:closedstrings}.
The states in this Hilbert space are labeled by collections of closed strings winding around the spatial circle, and the dynamics consists of local interactions that cut and reglue strings at the same point.
The sphere partition function can also be interpreted as a closed string amplitude, by foliating the worldsheet as in the middle diagram of figure \ref{figure:SU}.

In order to describe entanglement in the string theory picture, we must also describe the Hilbert space of an interval, and the entangling product that connects two intervals.
In section \ref{section:openstrings} we show how to describe the Hilbert space of an interval in terms of a canonical theory of \emph{open strings}.
Each open string carries a pair of Chan-Paton indices $i,j = 1,\ldots, N$ associated with its two endpoints.
The entangling product can be seen as a relation between the closed string and open string Hilbert spaces:
\begin{equation}
\mathcal{H}_\text{closed} \subset \mathcal{H}_\text{open} \otimes \mathcal{H}_\text{open}.
\end{equation}
This says that any closed string may be cut into open strings, but that not every collection of open strings can be reassembled into closed strings; this requires a specific matching of the Chan-Paton indices, which leads to a special entanglement structure of the closed string states.
We also derive the modular Hamiltonian for an interval on the sphere, and show that it corresponds to a Nambu-Goto term plus local interactions of the open strings.

The key point is that in the open string description, the partition function contains a sum over Chan-Paton indices associated with the string endpoints.
This results in an extra statistical weight associated to the $N^2$ possible values of the two Chan-Paton indices on each open string: this factor is directly responsible for the leading-order $N^2$ scaling of the entanglement entropy.
We find that the statistical weight associated with the open string edge modes is encoded in the sum over worldsheets by the $\Omega$-point singularity.
We argue that the two $\Omega$-points appearing on the sphere should be identified with the entangling surface, on which open strings are allowed to end.
This is similar to the way that open strings can end on a D-brane, so we call the resulting objects E-branes.
The E-brane encodes how the edge degrees of freedom appear at the level of the partition function, and also explains the appearance of Gross and Taylor's $\Omega$-point.
This is the central result of our paper.

In the large-$N$ limit the theory is described by two distinct chiral sectors, and in section \ref{section:openstrings} we focus for simplicity on the theory restricted to a single sector.
In section \ref{section:coupled} we describe the additional features the theory acquires when both sectors are included.
In this case the path integral contains worldsheets of two distinct orientations, and a new class of singularities --- orientation-reversing tubes --- can appear at the $\Omega$-points.
We show that these singularities, like the other aspects of the $\Omega$-points, are purely kinematical and simply enforce the unitarity condition $U^{-1} = U^\dag$.

We conclude in \ref{section:discussion} with some areas of future work, as well as a discussion on entanglement in string theory in higher dimensions.

\section{Two-dimensional Yang-Mills theory as a closed string theory}
\label{section:closedstrings}

Here we review the description of two-dimensional Yang-Mills theory as a theory of closed strings.
This will serve to establish some basic definitions for use later on, and motivates our generalization to open strings.
See ref.~\cite{Baez:1993gm} for an introduction, or ref.~\cite{Cordes:1994fc} for a more comprehensive review.

\subsection{The closed string Hilbert space}
\label{subsection:closedstringH}

We consider two-dimensional Yang-Mills theory with gauge group $G = \U(N)$, with Euclidean action
\begin{equation}
I = \frac{1}{4 g_\text{YM}^2} \int_M \sqrt{g} \tr[ F^{\mu \nu} F_{\mu \nu}].
\end{equation}
We first consider the Hamiltonian formulation of the theory where space is a circle of circumference $L$.
The theory has no local degrees of freedom, and the gauge-invariant variables can be constructed from the holonomy $U$ around the circle, and the nonabelian electric field $E^a$:
\begin{align}
U = \mathcal{P} \exp \left( i\int_0^L dx \, A_x(x) \right) \in G, \qquad \qquad E^a(x) = -i g_\text{YM}^2 \frac{\delta}{\delta A_a(x)} \in \mathfrak{g}.
\end{align}
The Hilbert space of states on the circle is the space of square-integrable class functions on the group manifold.
These are functions $\psi$ of the holonomy $U$ which are square integrable in the Haar measure, and invariant under the gauge transformation: $\psi(U) = \psi(g U g^{-1})$.

The Hamiltonian is proportional to the square of the electric field (there are no magnetic fields in 1+1 dimensions) which can be written explicitly as a second-order differential operator (the Laplacian) on the group manifold:
\begin{align}\label{H}
H
&= \frac{1}{2 g_\text{YM}^2} \int_0^L dx \; E^a(x) E^a(x) \nn
&= - \frac{g_\text{YM}^2}{2} \int_{0}^{L} dx \; \frac{\delta}{\delta A_a(x)} \frac{\delta}{\delta A_a(x)} \nn
&=\frac{ \lambda L}{2} \left(\sum_{ij} U_{ij} \frac{\partial}{\partial U_{ij}} +  \frac{1}{N} \sum_{ijkl} U_{ik} U_{jl} \frac{\partial}{\partial U_{jk}} \frac{\partial}{\partial U_{il}} \right)
\end{align}
where we have used the `t Hooft coupling $\lambda = g_\text{YM}^2 N$.

A natural basis of the Hilbert space is given by states $\ket{R}$ whose wave functions are the irreducible characters of the group:
\begin{equation}\label{char}
\braket{U|R} = \chi_R(U).
\end{equation}
The virtue of this basis is that it diagonalizes the Hamiltonian \eqref{H}.
The corresponding eigenvalues are proportional to the quadratic Casimir of each representation $H \ket{R} = \frac{\lambda L}{2} C_2(R) \ket{R}$.

To describe this theory as a string theory, we introduce a basis of states describing closed strings \cite{Baez:1994gk}.
Let $\sigma \in S_n$ be a permutation with $n_{l}$ cycles of length $l$. We define a state $\ket{\sigma}$ with $n_{l}$ closed strings winding $l$ times around the circle by 
\begin{align}\label{CS} 
\braket{U|\sigma} &=  \prod_{l=1}^{\infty} (\text{Tr} \,\,U^{l})^{n_{l}}, &  \sum_{l}n_{l} &=n.
\end{align}
Consistent with closed string indistinguishability,  these wave functions depend only on $\{ n_{l} \}$, which specifies the conjugacy class of $\sigma$.\footnote{In the full $\U(N)$ theory, the string states \eqref{CS} span only a subspace of $\mathcal{H}$.
These can be extended to a complete set by including wave functions which contain an additional factor of $\text{det}(U)^{m}$. 
In this paper we will restrict to the $m=0$ sector, for which the string states \eqref{CS} give a complete description.
The quantum number $m$ is related to the $\U(1)$ charge $Q$ labeling the total number of boxes in the Young diagram of $R$ by $Q = mN+n$.}

\begin{figure}
\centering
\includegraphics[scale=1]{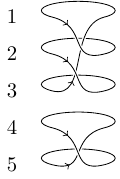}
\caption{The closed string configuration corresponding to the permutation $\sigma: 1\ra 2,2\ra 3,3\ra1,4\ra5,5\ra4$. The cycle lengths $(3,2)$ correspond to the winding numbers of the closed strings.}
\label{figure:CS}.
\end{figure}

The two sets of states are related by the Frobenius relations:
\begin{equation}\label{Fr}
\ket{R} = \sum_{\sigma \in S_n} \frac{\chi_R(\sigma)}{n!} \ket{\sigma}.
\end{equation}
Here we have used the fact that the irreducible representations of $\U(N)$ and of the permutation group $S_n$ can be represented by Young diagrams: $n$ is the total number of boxes in the Young diagram associated to $R$, and $\chi_R(\sigma)$ is the associated character of the permutation group. 
If we number the matrices $U$ in the product \eqref{CS} and represent each as an open string wrapped around the spatial circle, then the permutation $\sigma$ specifies the index contractions that glues the open string endpoints into a closed string configuration as illustrated in figure \ref{figure:CS}.
Applied to a closed string state $\ket{\sigma}$, the leading term of the Hamiltonian \eqref{H} counts the total winding number $n$, while the subleading term is a sum of interactions that locally cut and reconnect the strings.
In terms of permutations, each of these interactions act as transpositions $p \in T_{2}$ taking $\ket{\sigma}\rightarrow\ket{p\sigma}$ which changes the individual winding numbers, but not the total winding number. 
We can decompose the Hamiltonian into a free and interacting part:
\begin{align} \label{H'}
H &= H_0+ 2 H_1, \nn
H_0 \ket{\sigma} &= \frac{\lambda L}{2} n \ket{\sigma}, \nn
H_1 \ket{\sigma} &=  \frac{\lambda L}{2 N} \sum_{p \in T_{2}}\ket{ p \sigma } .
\end{align}
Since $H_1$ contains an explicit factor of $1/N$, the interaction can be treated as subleading in the large $N$ limit.
This is the limit in which the string theory becomes weakly coupled.

We now consider how to describe the Hilbert space of the theory in the large $N$ limit. 
The basis states $\ket{R}$ are labelled by representations, but we must define how to fix a representation as $N \rightarrow \infty$.
Naively one could fix the Young diagram for $\ket{R}$, thus fixing the right hand side of the Frobenius relation \eqref{Fr}, which remains well-defined in the large-$N$ limit.  
However, this keeps only states whose Young diagrams have a finite number of boxes, essentially truncating the theory in half: for example, this procedure would exclude the antifundamental representation, whose Young diagram is a column of height $N-1$.
The full Hilbert space at large $N$ can be identified as a subspace of the tensor product
\bea
\lim_{N \rightarrow \infty} \mathcal{H}_{\U(N)} \subset \mathcal{H}^+ \otimes \mathcal{H}^- ,
\eea
where  $\mathcal{H}^{\pm}$ are spanned by representations built out of finite numbers of fundamentals or antifundamentals respectively.
These two chiral sectors consist of closed strings winding in opposite orientations:
$\mathcal{H}^+$ is spanned by the closed string basis \eqref{CS} and $\mathcal{H}^-$ is spanned by a similar basis with $U$ replaced by $U^{\dagger}$.
This is the correct Hilbert space in the sense that the asymptotic $\frac{1}{N}$ expansion of the Yang-Mills partition function \eqref{ZYM} requires a sum over both chiral sectors.  
To make $\mathcal{H}^+$ explicit as a multi-string Fock space, we define the vacuum and the creation operator for a string winding $l$ times:
\begin{align}
\braket{U|0} &= 1\nn
\braket{U|a^{\dagger}_{l}|0} &= \text{Tr}(U^{l})\nn
\ket{\sigma} &= \prod_{l} (a^{\dagger}_{l})^{n_{l}} \ket{0} 
\end{align}
 $a_{l}^{\dagger}$ acts by multiplication by $S_{l}=\text{Tr}(U^{l})$. It can be shown that the corresponding adjoint operator is $a_{l} =  l  \frac{\partial} {\partial S_{l} }$,  which satisfies the commutation relations and normalizations:
\begin{align}
 [a^{\phantom{\dagger}}_{l},a^{\dagger}_{l'}] &= l \, \delta_{l l'} \nn
 \braket{\sigma|\sigma} &=  \prod_{l} l^{n_{l} } \, n_{l} ! \label{norm} 
 \end{align}
Note that the normalization of $a_l$ differs from the usual quantum-mechanical convention.  
The normalization of $\ket{\sigma}$ counts the number of permutations commuting with $\sigma$, which coincides with the order of the stabilizer subgroup, $\braket{\sigma|\sigma} = |C_\sigma|$, where
\begin{equation}
C_\sigma := \{ \tau : \sigma = \tau \sigma \tau^{-1} \}
\end{equation}
Let $T_\sigma$ denote the orbit of the permutation $\sigma$ under conjugation, 
\begin{equation}
T_\sigma := \{ \tau \sigma \tau^{-1} : \tau \in S_n \}.
\end{equation}
The orbit-stabilizer theorem states that the size of the orbit times the size of the stabilizer equals the order of the group:
\begin{equation}
|T_\sigma| \, |C_\sigma| = |S_n| = n!.
\end{equation}
Accounting for this normalization and closed string indistinguishability, we can write a resolution of identity in the $\ket{\sigma}$ basis,
\begin{align} \label{Res}
1 = \sum_{n} \sum_{\sigma \in S_{n}} \frac {\ketbra{\sigma}{\sigma}}{|T_\sigma| |C_\sigma|} = \sum_{n} \frac{1}{n!} \sum_{\sigma \in S_{n}} \ketbra{\sigma}{\sigma}
\end{align}
where in the first equality we have divided by $|T_\sigma|$ because all elements in the same conjugacy class represent the same closed string state, and divided by the normalization factor $|C_\sigma|$.
In the second equality, we applied the orbit-stabilizer theorem.

In terms of these closed string creation and annihilation operators, the Hamiltonian \eqref{H'} is  \cite{Minahan:1993np} 
\bea \label{H''}
H = \frac{\lambda L }{2} \left[ \sum_{k=1}^{\infty} a_{k}^{\dagger} a^{\phantom{\dagger}}_{k} +\frac{1}{N} \sum_{k,l =1}^{\infty} \left( a_{k+l}^{\dagger}a^{\phantom{\dagger}}_{k}a^{\phantom{\dagger}}_{l} + a_{k}^{\dagger}a_{l}^{\dagger}a^{\phantom{\dagger}}_{k+l} \right) \right] .
\eea
This Hamiltonian defines the closed string field theory dual to the chiral sector of two-dimensional Yang-Mills.
The first term is the free term, which is proportional to the length of the interval times total winding number of the strings.
This is the string tension, which is proportional to the total length of strings.
The second term is an interaction in which closed strings interact by splitting and joining via a cubic vertex.
This interaction corresponds to two strings of winding numbers $k$ and $l$ merging into a string of winding number $k + l$, and the reverse process.
This interaction preserves the total winding number, so it commutes with the free term.

The Hamiltonian \eqref{H''} does not capture the full theory, since the operators $a_k^\dag$ only creates states that can be obtained by acting on the vacuum state with multiplication by $\tr(U^k)$, which are all holomorphic wavefunctions.
To get the full $\U(N)$ theory we would have to include a sector of antiholomorphic wavefunctions. 
These can be thought of as strings winding in the opposite direction, and will be developed further in section \ref{section:coupled}.

Finally, at finite $N$, there are Mandelstam identities that force us to identify certain string states; for example, in the $\U(1)$ theory, $a_2^\dag \ket{0} = a_1^\dag a_1^\dag \ket{0}$.
These can be implemented by a projection operator as shown in ref.~\cite{Baez:1994gk}.
The effect of this projection is nonperturbative in the $1/N$ expansion, so we will neglect it in the following.

\subsection{Torus partition function}
To illustrate how the worldsheet expansion of two-dimensional Yang-Mills theory emerges, we now perform a perturbative expansion of the Yang-Mills partition function $Z_{T^{2}}$ on a torus and construct the corresponding closed string Feynman diagrams. 
We will see that the Hamiltonian evolution in Euclidean time $\beta$ of a multiply-wound closed string state traces out a multi-sheeted Riemann surface on which singularities appear whenever the strings interact by joining or breaking apart. 
Since the total winding number is conserved, the Hilbert space naturally divides into sectors labelled by $n$, which we denote $\mathcal{H}_n$, and the partition function decomposes as a sum:
\begin{equation}
Z_{T^2} = \tr ( e^{-\beta H} ) = \sum_{n} e^{- \frac{\lambda n A}{2}} \tr_{\mathcal{H}_n} (e^{-\beta H_1}).
\end{equation}
We recognize the leading term as the Nambu-Goto action of a string world sheet wrapping $n$ times around the torus of area $A = L \beta$.

To calculate the effect of the interaction term, we expand the exponential
\begin{align}
\tr_{\mathcal{H}_n} (e^{-\beta H_1}) 
&= \sum_{i}  \frac{(-1)^i}{i!}  \left(\frac{\lambda A}{N} \right)^i \tr_{\mathcal{H}_n} (H_1^i) \nn
&= \sum_{i}  \frac{(-1)^i}{i!}  \left(\frac{\lambda A}{N} \right)^i \sum_{\sigma \in S_n} \frac{1}{n!}\sum_{p_1,\ldots, p_i \in T_2} \bra{\sigma} p_1 \cdots p_i \ket{\sigma}. \label{trHn}
\end{align}
We recognize this as a sum over $i$ interaction points, with a factor of $1/i!$ indicating the indistinguishability of the interactions.\footnote{We are not aware of any physical interpretation for the factor of $(-1)^i$, but it is vaguely suggestive of a fermionic nature of the interaction points.}
The factor of $(1/N)$ is the string coupling associated with each interaction and  $\lambda A$ is a modulus factor, which comes from integrating over all possible locations for the interaction.
We can further expand this sum, by noting that the matrix element $\bra{\sigma} p_1 \cdots p_i \ket{\sigma}$ is nonzero precisely when there exists $\tau \in S_n$ such that
\begin{align}  \label{ps}
p \sigma = \tau \sigma \tau^{-1} 
\end{align}
where we have defined $p = p_1 \cdots p_i$.
Thus we can introduce an explicit sum over $\tau$, at the expense of dividing by the order of the stabilizer group, and rewrite the matrix element in \eqref{trHn} as
\begin{equation} \label{sps}
\braket{\sigma|p|\sigma} = \sum_{\tau \in S_n }  \delta(p\sigma, \tau \sigma \tau^{-1}).
\end{equation}
Above, the normalization of $\ket{\sigma}$ has canceled the division by $|C_\sigma|$. 
Using \eqref{trHn} and \eqref{sps}, the torus partition function is:
\begin{align}\label{ZT}
Z_{T^{2}}&= \sum_{n} \frac{e^{\frac{-n \lambda A}{2}}}{n!}  \sum_{i}  \frac{(-1)^{i}}{i!} \left( \frac{\lambda A}{N} \right)^{i}
\sum_{p_{1}... p_{i} \in T_{2} }   \sum_{\sigma, \tau \in  S_{n} } \delta(p\sigma, \tau \sigma \tau^{-1}).
\end{align}
\begin{figure}
\centering
\includegraphics[height=6cm]{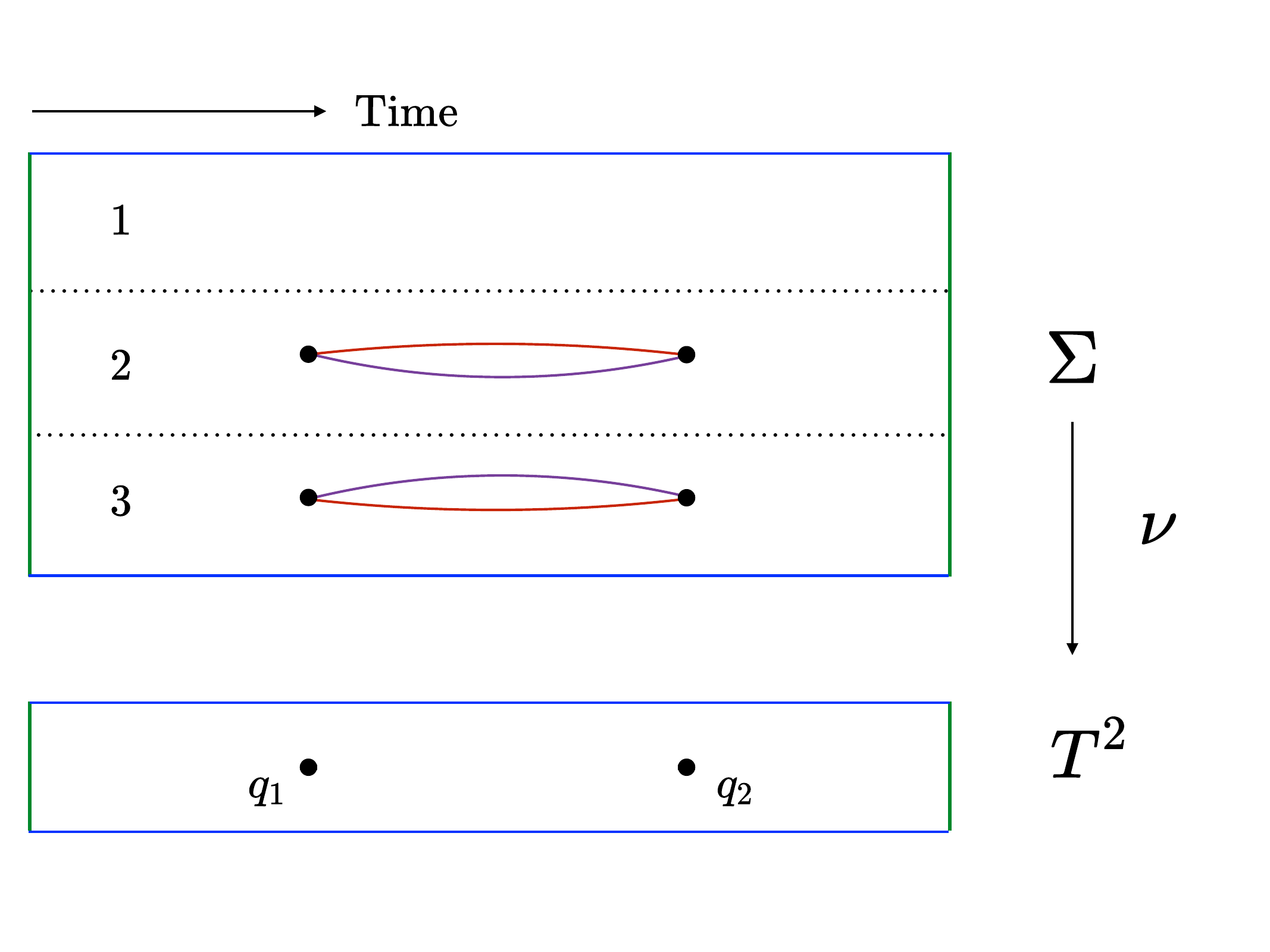}
\caption{A 3-sheeted covering map of the torus $T^{2}$ with two interaction branch points $q_{1},q_{2}$.   The cover map $\nu$ is defined here by vertical projection. A counterclockwise loop encircling the target space branch point $q_1$ lifts to a permutation $p_{1}:1\rightarrow 1,  2\rightarrow 3,3\rightarrow 2$.  A vertical time slicing of $\Sigma$ shows an initial closed string breaking into two and then joining back together again.} 
\label{cm}
\end{figure}

The expression \eqref{ZT} was obtained by Gross and Taylor \cite{Gross:1993hu}, who interpreted it as a sum over branched coverings of the torus $T^{2}$.   An $n$-sheeted covering $\nu : \Sigma \ra T^{2} $ with $i$ branch points $q_{1}, \ldots, q_{i} \in T^{2}$ is uniquely specified by a homomorphism
\bea
H_{\nu} : \pi_{1} (T^{2}  \setminus \{ q_{1}, \ldots , q_{i} \}) \ra S_{n} .
\eea This homomorphism describes how the $n$ sheets are permuted as we follow loops on $\Sigma$ obtained by lifting non-contractible curves on the punctured torus $T^{2} \setminus \{ q_{1},\ldots,q_{i} \}$.  
In particular each set of permutations $p_{1}, \cdots p_{i}  ,\tau,\sigma \in S_{n}$ satisfying the delta function constraint in \eqref{ZT} specifies such a homomorphism by describing how the $n$ sheets are shuffled when we encircle the branch points $q_{1},\cdots q_{i}$ and the two basis cycles of the torus.  
     
     As alluded previously, we interpret the covering space $\Sigma$ as a closed string worldsheet wrapping the torus $n$ times (see figure \ref{cm}) . Choosing fixed time slices running perpendicular to the branch cuts on $\Sigma$ shows that the branch points corresponds to interactions where closed strings break apart and join together. 
The $N$ dependence of the expression \eqref{ZT} can be understood from the Riemann-Hurwitz formula, which relates the Euler characteristic of the covering space $\chi(\Sigma)$ to the Euler characteristic of the target space $\chi(M)$ in the presence of $i$ elementary branch points.
\bea \label{RH}
\chi(\Sigma)= n  \chi(M) - i 
\eea 
Choosing $M=T^2$ gives $\chi(\Sigma)= - i $, so the $N$ dependence in \eqref{ZT} is $N^{\chi(\Sigma)}$, giving the correct exponent for the string coupling.  Finally, the $\frac{1}{n!}$ factor accounts for the redundancy from summing over homomorphisms $H_{\nu}$ that differ only by the relabelling of the $n$ sheets.  The cancellation is not exact, because there are relabelling of the $n$ sheets that fix $\nu$, leading to a symmetry factor that we denote by $\frac{1}{|S_{\nu}|}$.  Thus we are led to a closed string partition function which can be expressed in a compact form \cite{Gross:1993hu} : 

\begin{align}\label{Zst} 
Z =  \int_{C(M)}  d\nu  \left( \frac{1}{N} \right)^{2g-2} \frac{ (-1)^i}{|S_{\nu}|}  e^{\frac{-n \lambda A}{2}}. 
\end{align}
Here $C(M)$ denotes a set of covering maps of $M$.

\subsection{Sphere partition function}

The partition function on a sphere is obtained by gluing together the path integrals on a twice-punctured sphere and two infinitesimal disks.  
In the string picture, we can view the gluing as a closed string evolution between initial and final states inserted at the two punctures, known as $\Omega$-points \cite{Gross:1993yt}:
\begin{equation} \label{EB}
Z^+_{S^2} = \bra{\Omega} e^{-\beta H} \ket{\Omega}.
\end{equation}
The superscript ``$+$'' denotes that we are considering only a single chiral sector, \textit{c.f.} section \ref{subsection:closedstringH}.
The state $\ket{\Omega}$, which is defined by the path integral over an infinitesimal disk, can be written in a number of different ways.

In the holonomy representation the state $\ket{\Omega}$ is given by
\begin{equation}\label{delta}
\braket{U|\Omega} = \delta(U),
\end{equation}
where here $\delta$ denotes the Dirac delta function in the Haar measure supported on the identity element.
This expresses the fact that as a circle contracts its holonomy becomes trivial.\footnote{
Note that unlike the case of a conformal theory, the state generated by the path integral on a small disk with no insertions is not the vacuum. The vacuum is a constant wavefunction in the $U$ basis, and corresponds to the path integral over a \emph{large} disk. 
Two-dimensional Yang-Mills theory is not scale-invariant, so a small disk is not equivalent to a large one.
}

This state can be written equivalently in the representation basis as
\begin{equation} \label{BS}
\ket{\Omega} = \sum_R \dim R \ket{R}.
\end{equation}
The factor of $\dim R$ has an important effect on the entanglement entropy: it leads to an additional $\log \dim R$ factor in the entropy that represents a sum over degenerate states in each representation of the gauge group \cite{Donnelly2014a}.

In the string basis, we can similarly write \footnote{To derive  \eqref{Omega} we used the fact that for $R$ corresponding to a Young diagram with $n$ boxes
\bea
 \dim R= \sum_{\sigma \in S_{n} } \frac{\chi_{R}(\sigma)}{n!} N^{K_\sigma} ,
\eea
as can be seen by setting $U=1$ in the Frobenius relation \eqref{Fr}.}
\begin{equation} \label{Omega}
\ket{\Omega} = \sum_n \frac{1}{n!} \sum_{\sigma \in S_n} N^{K_\sigma} \ket{\sigma}
\end{equation}
where $K_\sigma$ denotes the number of cycles in the permutation $\sigma$.
The amplitude $\braket{\sigma|\Omega}=N^{K_\sigma}$ plays a similar role to the $\dim R$ factor in the Yang-Mills description: it leads to a counting of degenerate states that contribute to the entropy, as we will show in section \ref{section:openstrings}.

\begin{figure}
\centering 
\includegraphics[height=4 cm]{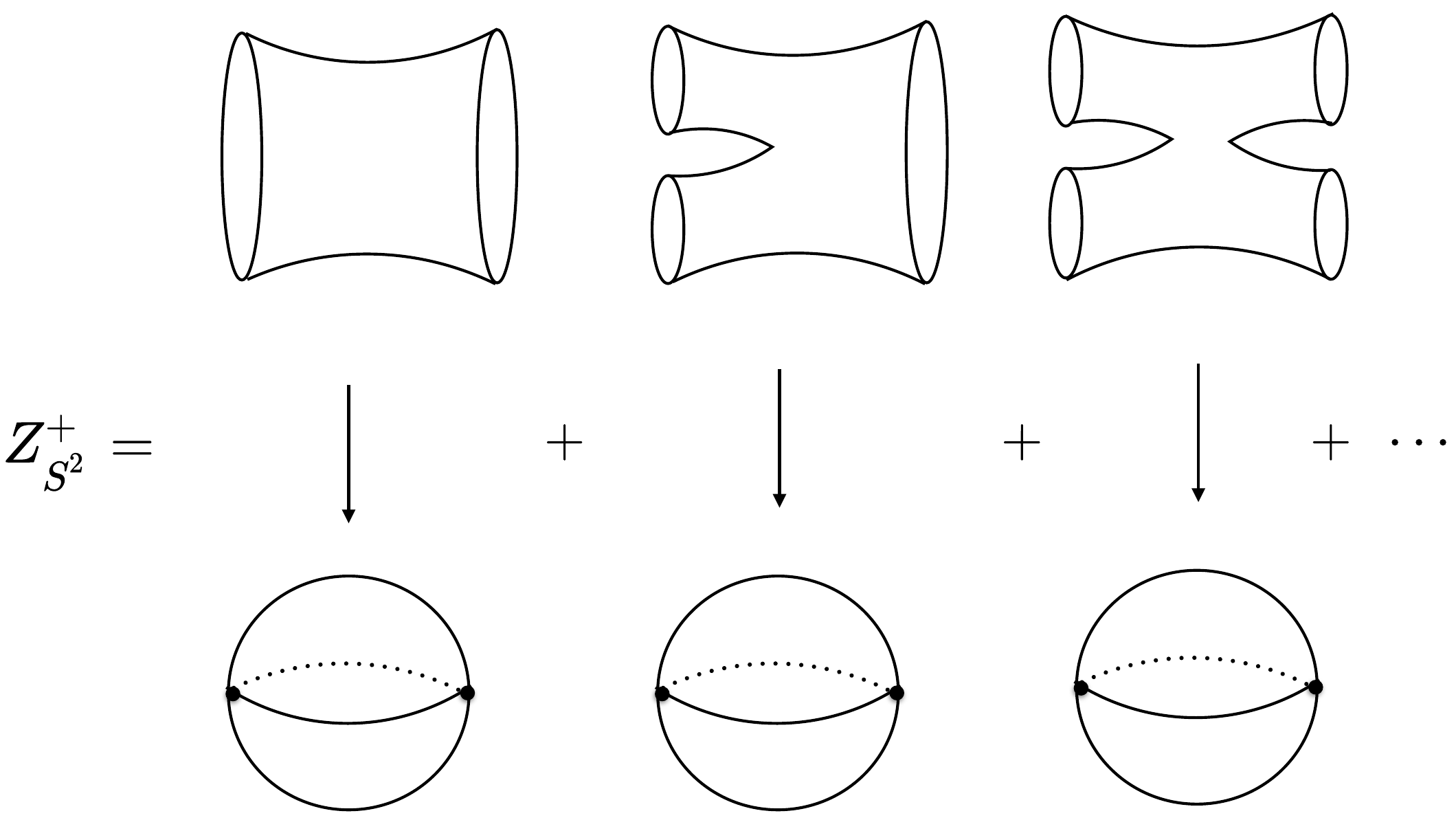}

\caption{$Z^{+}_{S^2}$ can be expressed as a sum over worldsheets with holes corresponding to closed strings that wind around the $\Omega$-points on $S^2$.}
\label{holes}
\end{figure}

Using the state \eqref{Omega} and the closed string Hamiltonian (\ref{H'}), we can then write the sphere partition function as
\begin{align} \label{ZS2'}
Z_{S^2} &= \bra{\Omega} e^{-\beta H} \ket{\Omega} \nn
&= \sum_{n} \frac{e^{-\frac{n \lambda A}{2}}}{n!} \sum_{i} \frac{(-1)^i}{i!} \left( \frac{\lambda A}{N} \right)^i \sum_{\sigma \in S_{n}} \sum_{p_1, \cdots, p_i \in T_2} N^{K_\sigma} N^{K_{p_1 \cdots p_i \sigma}},
\end{align}

We interpret the sum over $\sigma \in S_{n}$ as a sum over world sheets $\Sigma$ on which an initial state of $K_{\sigma} $ closed strings evolves into a final state of $K_{p _1 \cdots p_i \sigma}$ closed strings via $i$ interactions (see figure \ref{holes}). 
Such a world sheet wraps the punctured sphere $n$ times with $i$ elementary branch points $p_{1}, \ldots, p_{i} $. 
The closed strings in the external states correspond to infinitesimal holes on the world sheet that wind around the two $\Omega$-points according to the cycle lengths of $\sigma$, and $p _1 \cdots p_i \sigma $.
Closing up the holes leads to a covering map for which the $\Omega$-point is a multiple branch point singularity labeled by the permutation $\sigma$ or $p_{1}, \cdots p_{i} \sigma$ (see figure \ref{holes}).
Thus, we can once again appeal to \eqref{RH} to determine the Euler characteristic 

\bea
\chi(\Sigma)=-i +  K_{\sigma} + K_{ p _1 \cdots p_i \sigma} ,
\eea
consistent with the power of the string coupling in (\ref{ZS2'}).
As before, the division by  $n!$ accounts for the partial redundancy due to relabelings of the $n$ sheets, and leads to the correct symmetry factor for each diagram.   

\begin{figure}
\centering 
\includegraphics[height=4 cm]{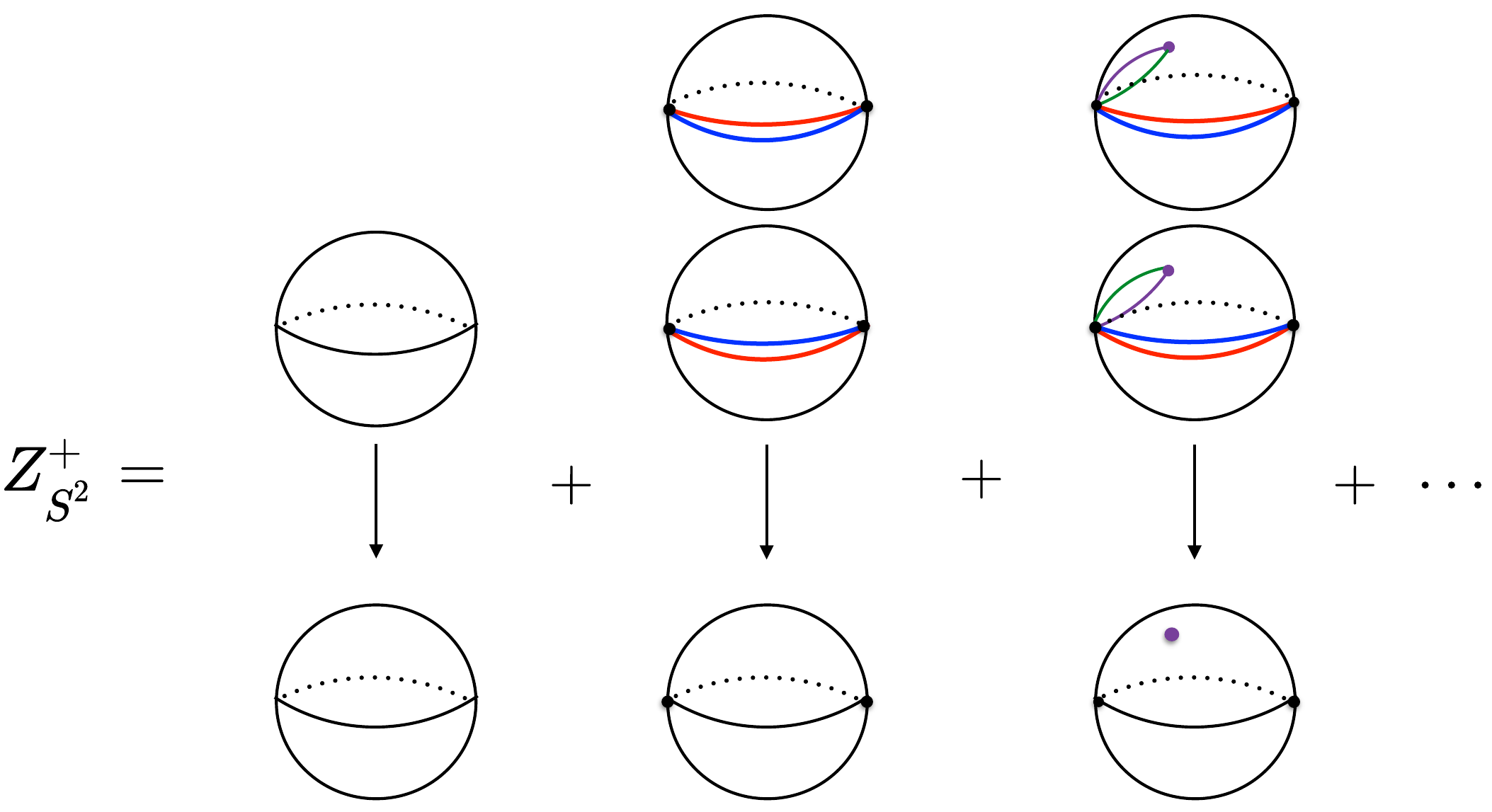}

\caption{ By closing up the boundaries of the worldsheet into branch points and introducing appropriate branch cuts, we can  present the worldsheet as a covering space of the sphere, with covering map represented by vertical projection.  The first is a single covering with no interaction and $\Omega$-point singularities. The second is a double cover of the sphere with two $\Omega$-points. The third term corresponds to the ``pair of pants" diagram, now presented as a double cover with an interaction branch point inserted, which is connected to an $\Omega$-point via a branch cut. }
\label{GT}
\end{figure}

Thus we conclude that the closed string expansion \eqref{Zst} continues to hold on the sphere, provided that the space of covering maps is extended to allow for two $\Omega$-points, which are fixed point of $M$ at which multiple branch points can appear.  
While the presence of such branch point singularities was derived in \cite{Gross:1993yt}, we have introduced a new interpretation of them as sources that emit and absorb closed strings with coupling $N=\frac{1}{g_{\text{string}} }$ per closed string. 
The state $\ket{\Omega}$ is therefore analogous to a D-brane boundary state, which has the same coupling to closed strings.
We will call it an \emph{entanglement brane} or E-brane.
The presence of brane-like objects suggests the presence of open strings which would couple to the E-brane.
In the next section we will pursue the open string description of the $\Omega$-points and reinterpret them as entangling surfaces in string theory.

\section{Angular quantization on the sphere and open strings}
\label{section:openstrings}

In the preceding section we have expressed the sphere partition function in terms of closed strings propagating between two E-brane boundary states $\ket{\Omega}$.
This sphere partition function is the key ingredient that enters in calculating the entanglement entropy in two-dimensional de Sitter space.
In order to see which states are being counted by this entanglement calculation, we have to express the partition function as a trace of the form $\tr e^{-\beta H_V}$ --- the key is to identify the appropriate Hilbert space and modular Hamiltonian $H_V$. 
This means we have to foliate the same sphere diagram by intervals as in the right diagram of figure \ref{figure:SU}.
These intervals are anchored at the two poles of the sphere, and it is natural to identify these two points with the two $\Omega$-points that appear in the sphere partition function.
Treating the angular coordinate $\phi$ as a Euclidean time variable we will see that the sphere partition function $Z_{S^2}$ naturally describes a canonical ensemble of opens strings at finite temperature.
We will further see that the modular Hamiltonian takes a similar form as in the closed string theory; it consists of a Nambu-Goto term plus local interactions that cut and reglue the open strings.

Note that at large $N$, the Gross-Taylor model contains two interacting chiral sectors.
For simplicity, we will initially treat a single chiral sector, and generalize to the coupled theory in section \ref{section:coupled}.
We will denote the partition function of a single chiral sector on the sphere as $Z_{S^2}^+$.

\begin{figure}
\centering
\includegraphics[height=4 cm]{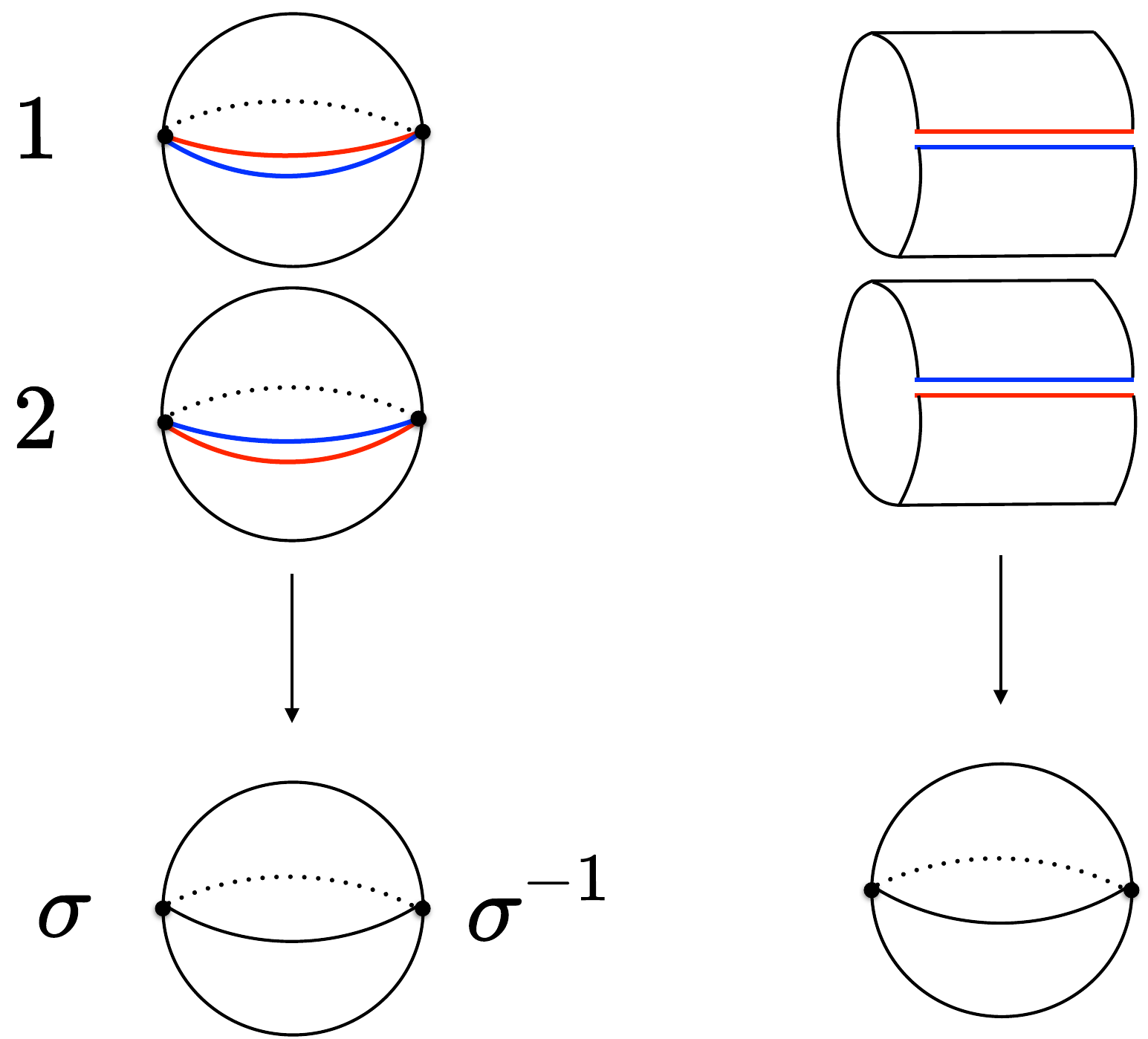}

\caption{
The figure on the left shows a term in the Gross-Taylor expansion corresponding to a two-sheeted branched covering of the sphere with two $\Omega$-point singularities.  
These two fixed points on the target sphere are labelled by the permutation $\sigma : 1 \rightarrow 2, 2 \rightarrow 1$, since circling these points interchanges the two sheets. 
On the right, we have opened up each branch point on the covering space into a \emph{single} connected boundary, representing an external closed string state given by $\ket{\sigma}$.  In the open string channel, this is interpreted as a finite temperature loop diagram in which an open string anchored on the $\Omega$-points winds twice around the sphere. }
\label{os2}
\end{figure}

\subsection{Entanglement of the Hartle-Hawking state}

\begin{figure}
\centering
\includegraphics[height=4 cm]{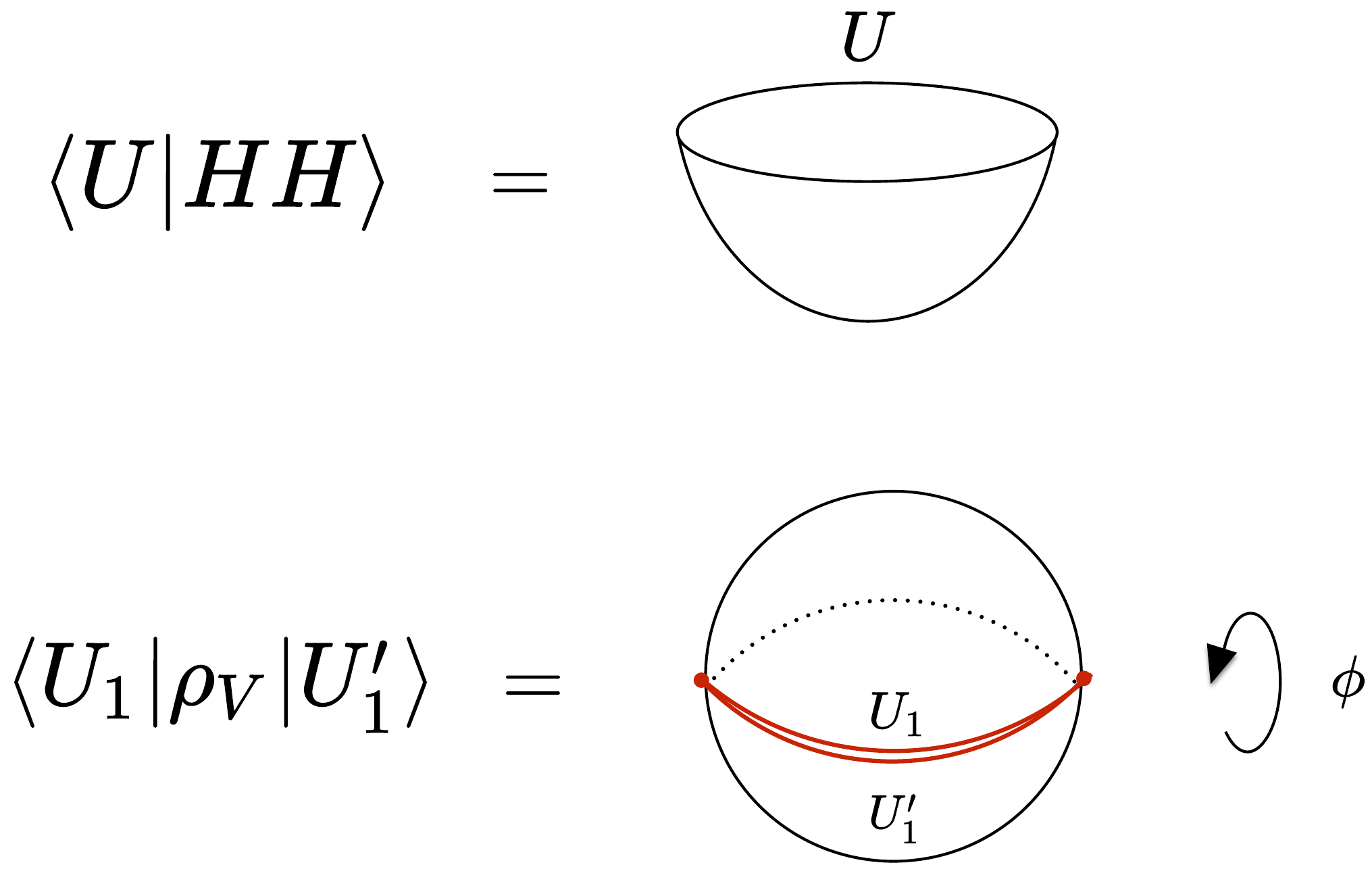}
\caption{
The Yang-Mills path integral on a hemisphere gives the unnormalized Hartle-Hawking wave function, and the path integral on the sphere computes the square of its norm.
Changing the periodicity of the angular coordinate $\phi$ to $2 \pi n$ yields $\tr(\rho_V^n)$.
}
\label{HH} 
\end{figure}

To set up our entanglement calculation we consider the Hartle-Hawking state $\ket{HH}$, which is prepared by the Euclidean path integral over the hemisphere geometry pictured in figure \ref{HH}.
This defines a state in the closed string Hilbert space associated with the circular boundary of the hemisphere.
Wick rotated to Lorentzian signature, this state corresponds to the de Sitter-invariant vacuum state, which reduces to a thermal state in the static patch.

We will divide the circle into a semicircle $V \subset S^1$ and its complement $\bar V$.
The entanglement between the degrees of freedom in $V$ and those in $\bar V$ is characterized by the reduced density matrix 
\begin{equation}
\rho_{V} = \tr_{\mathcal{H}_{\bar V}} \ketbra{HH}{HH}.
\end{equation} 
The unnormalized density matrix $\rho_V$ can be expressed as a path integral on the sphere with a cut along $V$ as in the right of figure \ref{HH}.
The sphere partition function $Z_{S^2}$ can be obtained by tracing over $V$, which corresponds to ``sewing up'' the cut in the sphere:
\bea \label{aq}
Z_{S^2} = \tr_{\mathcal{H}_V}  \rho_{V} = \tr_{\mathcal{H}_V}  e^{-H_V}.
\eea
Thus we can regard $Z_{S^2}$ as a thermal partition function respect to the modular Hamiltonian $H_{V} := -\log \rho_{V}$, which generates Euclidean time evolution in the angular direction orthogonal to $V$.
The entanglement entropy is simply the thermal entropy of this ensemble at inverse temperature $\beta = 1$.

Following the same steps that lead to the derivation of the Hamiltonian on the circle \eqref{H}, one finds that the modular Hamiltonian $H_{V}$ is simply the quadratic Casimir operator \eqref{C2} acting on the Hilbert space $\mathcal{H}_V$.
This is most easily proven when the size of the interval $V$ is exactly half the size of the great circle. 
In this case, the vector field $\xi^{a}$ that generates flow in entanglement time corresponds to a rigid rotation of the sphere, and the modular Hamiltonian of the Hartle-Hawking vacuum is the (suitably normalized) Killing energy which generates this symmetry \cite{Jacobson:1994fp}:
\begin{align}
H_{V} &= 2\pi    \int^{\pi}_{0} d\theta \sqrt{q} \, T_{ab} \xi^{a} n^{b} .
\end{align}
To compute $H_{V}$ explicitly we use spherical coordinates $ ds^{2}=r^{2} (d\theta^{2} + \sin^{2}(\theta)d\phi^{2})$ and choose $V$ to be the line segment $\phi = 0$, $\theta \in [0,\pi]$.
The entangling surface then consists of the two poles $\theta = 0$ and $\theta = \pi$.
The Killing vector that fixes the entangling surface is $\xi =\frac{\partial}{\partial \phi} $ the unit normal is $n = \frac{1}{r \sin(\theta)} \frac{\partial}{\partial \phi}$ and $\sqrt{q} = r$ is the volume element on a slice of fixed $\phi$.
The Yang-Mills energy density is $T^\phi{}_\phi=\frac{g_\text{YM}^{2}}{2} \tr(E^2) = \frac{\lambda}{2N} C_2$, which is constant over the interval, so we find:
\begin{align} \label{HV}
H_V = 2 \pi \int_0^\pi d \theta \, r^2 \sin(\theta) \frac{g_\text{YM}^2}{2} \tr(E^2) = 2 \pi r^2 \frac{\lambda}{N} C_2 = \frac{\lambda A}{2N} C_2
\end{align}
where we have used $A = 4 \pi r^2$ for the sphere.
Note that the modular Hamiltonian \eqref{HV} is the same for an interval of any size on the sphere; by area-preserving diffeomorphism symmetry any two such intervals are equivalent, since the theory is only sensitive to the topology and total area.

The entanglement entropy is obtained by taking the derivative of the partition function
\begin{equation}
S = \left. (1 - \beta \partial_\beta) \log Z_{S^2} \right|_{\beta = 1}.
\end{equation}
This corresponds to a deformation of the background that introduces a small conical singularity at the entangling surface, taking the angular period from $2 \pi$ to $2 \pi \beta$. 
Since 2D Yang-Mills theory is only sensitive to the area and Euler characteristic $\chi$, this variation corresponds to varying the area linearly with $\beta$ while keeping $\chi$ fixed.
Thus the entropy of the Hartle-Hawking state is given simply by:
\begin{equation}
S = (1 - A \partial_A) \log Z.
\end{equation}
For more general situations (higher genus surfaces and larger numbers of intervals) the entropy involves analytically continuing the partition function in the Euler characteristic; we will not consider those cases here.\footnote{In these cases one cannot find a foliation of the manifold that degenerates at the entangling surface as in the case of a single interval on $S^2$. Thus we expect to find a nonlocal modular Hamiltonian.}

Before giving the string description of the Hilbert space and modular Hamiltonian, we can describe them in terms of gauge theory variables.
There the Hilbert space of an interval can be expressed in terms of states $\ket{R,a,b}$ where $R$ is an irreducible representation of $\U(N)$ and $a,b$ are indices in that representation. 
This basis diagonalizes the Casimir operator $C_2$, whose eigenvalues, which we denote we denote $C_2(R)$, depend only on the choice of representation.
This leads to the expression for $Z_{S^2}$ in the representation basis:
\bea
Z_{S^2}=\sum_{R} (\dim R)^2e^{-\frac{\lambda A}{2N} C_2(R)}.
\eea 
The additional factor of $(\dim R)^2$ comes from counting the edge modes of the gauge theory, which consist of one additional degree of freedom at each endpoint transforming in the representation $R$.

We will see that there is an analogous stringy interpretation of these edge modes.
The Hilbert space of an interval can be described in terms of open strings.
In the string picture, the resolution of each $\Omega$-point results in a set of small disks cut out of the worldsheet, each with a factor of $N$ associated with the sum over Chan-Paton factors.
This is precisely what would be obtained by placing $N$ E-branes at the regulated entangling surface, and allowing open strings to end there.
The edge modes in the string theory description are simply the Chan-Paton factors of the open strings.

We will also give an open string description of the modular Hamiltonian $H_{V}$ in which the closed string coupling $N=\frac{1}{g_{\text{string}}}$ to the $\Omega$-point in \eqref{ZS2'} arises from summing over $N$ Chan-Paton factors associated to open string endpoints anchored on the entangling surface.
To understand the origin of these edge modes we proceed by deriving the open string description of the Hilbert space on the interval $V$.

\subsection{The open string Hilbert space}
\label{os}

In order to describe entanglement of string states, we first need a stringy description of the Hilbert space of an interval.
This Hilbert space is the space of square-integrable functions on the group manifold, $L^2(G)$.
Unlike the states on a circle, the states on an interval are not required to be class functions.
Here we describe a class of open string states analogous to the closed string states of section \ref{section:closedstrings}.

\begin{figure}
\centering 
\includegraphics[scale=1]{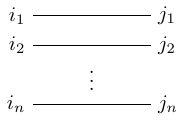}
\caption{The state $\ket{IJ}$ represents a configuration of $n$ open strings with Chan-Paton factors $(i_1, j_1) \ldots (i_n, j_n)$.}
\label{openstrings}
\end{figure}

Analogous to the closed string states of a circle, we introduce the following open string states of an interval.
Consider a state of $n$ open strings, each carrying Chan-Paton factors $i,j = 1 \ldots N$.
Letting $I = (i_1, \ldots, i_n)$ and $J = (j_1, \ldots, j_n)$, we define the state $\ket{I,J}$ by the wave functional
\begin{equation} \label{openstringstate}
\braket{U| I,J} = U_{i_1 j_1} U_{i_2 j_2} \cdots U_{i_n j_n}.
\end{equation}
Note that the two Chan-Paton indices $i$ and $j$ of a string are distinguished, because one transforms in the fundamental representation, and the other transforms in the antifundamental. 
In other words, the open strings are oriented.  
We depict such an open string state in figure \ref{openstrings}.

Note that unlike closed string states, open string states with different labels $I,J$ are not orthogonal.
There are two independent reasons for this.
The first is that open strings are indistinguishable, so $\ket{IJ}$ and $\ket{\sigma(I) \sigma(J)}$ label the same state, when $\sigma$ is any permutation.
This overcompleteness can be accounted for by labelling each such state with occupation numbers $\{ n_{ij} \}_{i,j=1}^n$ which count the number of strings with Chan-Paton factors $(i,j)$.
However there is a further non-orthogonality: the state $\ket{IJ}$ has nontrivial overlap with $\ket{I \, \sigma(J)}$.
This is a consequence of the nontrivial inner product on the space $L^2(G)$.

The Hilbert space of an open string is $L^2(\U(N),dU)$ where $dU$ is the Haar measure.
We can use the matrix elements $U_{ij}$ as coordinates, and the Haar measure is given by:
\begin{equation}
dU = \frac{\mathcal{N}}{\det(U)^N} \prod_{i,j} d U_{ij}.
\end{equation}
We can check that this is invariant under left and right multiplication, which uniquely determines the Haar measure.\footnote{Up to the normalization factor $\mathcal{N}$ which plays no role in our discussion}

The multiplication operator $U_{ij}$ can be viewed as a creation operator that creates an open string with Chan-Paton indices $(i,j)$.
We can also consider the operator $\frac{\partial}{\partial U_{ij}}$ which annihilates an open string and satisfies $[\partial_{ij}, U_{kl}] = \delta_{ik} \delta_{jl}$.
Unlike the usual ladder operators, the annihilation operator $\partial_{ij}$ is not the adjoint of the creation operator $U_{ij}$.
To find its adjoint, we have to consider the inner product
\begin{align}
\int dU f(U)^* \frac{\partial}{\partial U_{ij}} g(U) 
&= - \int \frac{\partial}{\partial U_{ij}} \left(\frac{\mathcal{N}}{\det(U)^N} f(U)^* \right) g(U) \prod_{k,l} dU_{kl} \nn
&= \int dU \left( N U^{-1}_{ji} f(U)^* - \frac{\partial}{\partial U_{ij}} f(U)^* \right) g(U) \nn
&= \int dU \left( N U_{ij} f(U) + U_{ik} U_{lj} \frac{\partial}{\partial U_{lk}} f(U) \right)^* g(U)
\end{align}
In the last line we have used the unitarity condition, which implies
\begin{equation} \label{partial}
\frac{\partial}{\partial U_{ij}} = - U^\dag_{ki} U_{jl}^\dag \frac{\partial}{\partial U^\dag_{kl}}.
\end{equation}
Thus we find that
\begin{equation} \label{partialdagger}
\left(\frac{\partial}{\partial U_{ij}} \right)^\dag = U_{ik} \frac{\partial}{\partial U_{lk}} U_{lj} = N U_{ij} + U_{ik} U_{lj} \frac{\partial}{\partial U_{lk}}.
\end{equation}
This suggests that to leading order in large $N$, we can treat the second term as subleading to the first, so that $U_{ij}$ and $\partial_{ij}^\dag$ act as rescaled creation and annihilation operators.
This is true as long as we consider states with a small number $n \ll N$ of strings, otherwise the second term receives an $n$-fold enhancement making its influence comparable to that of the first term.
This approximation is discussed further in \ref{subsection:freestring}.

\subsection{Electric fields and quadratic Casimir}

Having described the Hilbert space of an interval in terms of open strings, we now give the open string description of the nonabelian electric field operators, and the quadratic Casimir which determines the modular Hamiltonian of an interval.

The Hilbert space of an interval carries two commuting actions of $\U(N)$, which are given by left and right multiplication.
Their generators are the nonabelian electric fields at the endpoints of the interval, which we call the left and right electric fields.

The left electric field is given by
\begin{equation}
E^L_{ij} = U_{ik} \frac{\partial}{\partial U_{jk}}.
\end{equation}
This generates the left action of $\U(N)$ on itself when contracted with an $\mathfrak{u}(N)$ generator.
The electric field satisfies $(E_{ij}^L)^\dag = E^L_{ji}$, and its commutation relations represent the Lie algebra $\mathfrak{u}(N)$:
\begin{equation}
[E^L_{ij}, E^L_{kl}] = \delta_{jk} E^L_{il} - \delta_{il} E^L_{kj}.
\end{equation}
Acting on the string states, the action of $E_{ij}$ is to transform strings with Chan-Paton indices $(j,k)$ into strings with Chan-Paton indices $(i,k)$.
The diagonal element $E^L_{ii}$ (without summation on $i$) counts the number of open strings with left Chan-Paton index $i$.

There is also an analogous generator of transformations on the right, 
\begin{equation}
E^R_{ij} = U_{ki} \frac{\partial}{\partial U_{kj}}.
\end{equation}
This satisfies the same algebra and adjoint relation as $E^L$.
Moreover, the left and right electric fields commute:
\begin{equation}
[E^L_{ij}, E^R_{kl}] = 0.
\end{equation}

Like in the closed string Hilbert space, we can define a quadratic Casimir operator $C_2$. Just as the Hamiltonian for evolution along a cylinder was proportional to $C_2$, the modular Hamiltonian for an interval on the sphere is also proportional to $C_2$.
The Casimir is given in terms of the electric field as 
\begin{equation} \label{C2}
C_2 = \tr(E^2) = \sum_{i,j} E^L_{ij} E^L_{ji} = \sum_{i,j} E^R_{ij} E^R_{ji} = N U_{ik} \frac{\partial}{\partial U_{ik}} + U_{ik} U_{jl} \frac{\partial}{\partial U_{jk}} \frac{\partial}{\partial U_{il}}.
\end{equation}
The Casimir operator commutes with both the left and right electric fields.

The Casimir operator \eqref{C2} naturally splits into a leading term and an interaction subleading in the $1/N$ expansion.
The leading term just counts the number of open strings. 
The effect of the subleading quartic interaction is to cut two open strings, and glue them back together in a different order: we take two open strings with Chan-Paton factors $(j,k)$ and $(i,l)$ and replace them with strings with Chan-Paton factors $(i,k)$ and $(j,l)$.
Note that the interaction preserves the number of strings, so the two terms commute.
We write this as
\begin{equation}\label{C2expansion}
C_2 = N n + 2 H_1,
\end{equation}
where the interaction term implements a transposition.
The factor of $2$ accounts for the double counting in the expression \eqref{C2expansion}:
for a fixed pair of strings the interaction term in \eqref{C2expansion} acts twice, once with the indices as $ij$ and once with them as $kl$.
Introducing the permutation operators $\sigma$ such that 
\begin{equation}
\sigma \ket{IJ} = \ket{I \, \sigma(J)}
\end{equation}
we can write $H_1$ as a sum over all transpositions:
\begin{equation}\label{Hint}
H_1 = \sum_{\sigma \in T_2} \sigma.
\end{equation}

Given a state $\ket{\sigma}$ in the closed string Hilbert space, we can view it as a state in the product of two open string Hilbert spaces as follows:
\begin{equation} \label{sigmadecomp}
\ket{\sigma}= \sum_{I J} \ket{I J} \ket {J \, \sigma(I)}.
\end{equation}
While any closed string state can be written as a state in the tensor product of two open string Hilbert spaces, the converse is not true.
This is because the states coming from closed strings have the further constraint that the number of states with left Chan-Paton index $i$ on one interval must equal the number of states with right Chan-Paton index $i$ on the other interval.
We can see that this is a very significant restriction, as the closed string Hilbert space (at fixed $n$) has a dimension of order $N^0$, whereas the dimension of the open string Hilbert space grows as $N^{2n}$.
In the Yang-Mills description the restriction to closed string states corresponds to matching of the nonabelian electric field across the entanglement cut.

\subsection{The open string partition function}

Now we are ready to derive the Gross-Taylor expression \eqref{ZS2'} for the sphere partition $Z^{+}_{S^{2}}$ in the open string channel.
The open string partition function is simply the thermal partition function of the modular Hamiltonian:
\begin{equation}\label{ZS2}
Z_{S_2}^+ = \tr \left( e^{-\frac{\lambda A}{2 N} C_2} \right)
\end{equation}
where the trace is over the open string Hilbert space, and $C_2$ is the open string Casimir operator.

\begin{figure}
\centering 
\includegraphics[scale=.25]{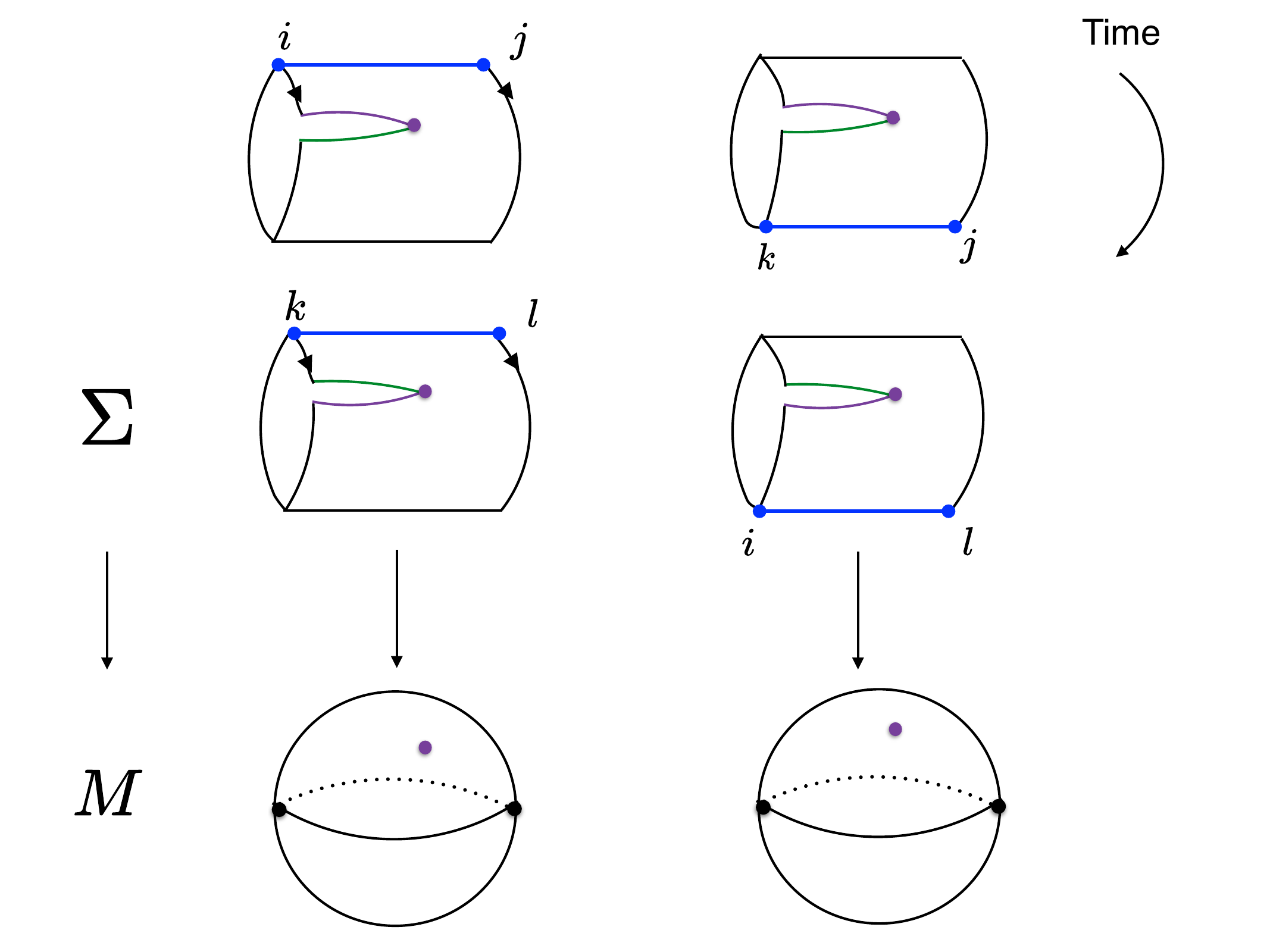}
\caption{Evolution under the open string modular Hamiltonian creates cylindrical worldsheets where time runs transverse to branch cuts.
The figure shows a pair of open strings and their Chan-Paton indices at two different time slices.  The states on the left endpoints are exchanged as they pass through the branch cut.}
\label{EH}
\end{figure}

Now we can do a perturbative expansion of the interaction term $H_{1}$ to obtain open string Feynman diagrams.  Since the interaction term commutes with the free term that counts the number of open strings, we can write
\begin{align}\label{ZS}
\tr(\rho_{V}) &= \sum_{n} e^{\frac{-n \lambda A}{2}} \sum_{i} \frac{(-1)^i}{i!}
\left( \frac{\lambda A }{ N} \right)^i   \tr_{\mathcal{H}_n} (H_{1}^i)
\end{align}
where here $\mathcal{H}_n$ is the sector of the Hilbert space with $n$ open strings.
Due to open string indistinguishability, we must be careful to count each state only once.
Let $T_{IJ}$ denote the orbit of the state $IJ$, $T_{IJ} = \{ \sigma(I) \sigma(J) : \sigma \in S_n \}$.
Since all elements of the orbit label equivalent open string states, we have to divide by the size of the orbit:
\begin{align}
\tr(\rho_{V}) &= \sum_{n} e^{\frac{-n \lambda A}{2}} \sum_{i}   \frac{(-1)^i}{i!} \left(\frac{\lambda A }{ N} \right)^{i}  \sum_{IJ} \frac{1}{|T_{IJ}|}  \langle IJ | H_{1}^{i} |IJ\rangle \nn
&= \sum_{n} e^{\frac{-n \lambda A}{2}} \sum_{i}  \frac{(-1)^i}{i!} \left(\frac{\lambda A}{N} \right)^{i}  \sum_{IJ}  \frac{1}{|T_{IJ}|} \sum_{p_{1},...p_{i} \in T_2}\braket{ IJ |p| IJ}. 
\end{align} 
In the last line we have written the interaction term as a sum of transpositions $p_1, \ldots, p_i$ and denoted $p = p_{1} \cdots p_{i}$.
The matrix element $\braket{IJ|p|IJ}$ is nonzero if and only if there is a permutation $\tau$ such that
\begin{align} \label{c}
I=\tau(I), \qquad p(J) = \tau(J)
\end{align}
For a given $I,J$, we denote the stabilizer subgroup as $C_{IJ} = \{ \tau \in S_n : \tau(I) = I, \tau(J) = J \}$. 
The number of permutations satisfying \eqref{c} is the order of the stabilizer subgroup $|C_{IJ}|$, since given any element $\tau$ satisfying the constraint, any other permutation $\sigma$ satisfying \eqref{c} must also belong to the coset $\tau C_{IJ}$, which has the same number of elements as $C_{IJ}$. 
Then using again the orbit-stabilizer theorem $n!=  |T_{IJ}| |C_{IJ}|$, the sum over open string states at fixed $n,i$ gives:  
\begin{align}
\sum_{I,J} \frac{1}{|T_{IJ}|} \braket{IJ | I p(J)} 
&= \sum_{IJ} \sum_{\sigma \in S_n} \frac{ \delta(I, \sigma(I)) \delta (J, \sigma p(J)) }{|T_{IJ}| |C_{IJ}|} \\
&= \frac{1}{n!} \sum_{\sigma \in S_n} N^{K_\sigma} N^{K_{p^{-1} \sigma}}.
\end{align}
Each term in this sum counts the number of states compatible with an $n$-sheeted open string worldsheet on which interaction branch points $p_{1}\dots p_{i} $ have been inserted in the bulk, opening branch cuts extending to the the boundary.  
As shown in figure \ref{EH}, such branch cuts implement the exchange of Chan-Paton indices produced by the open string modular Hamiltonian.
$K_{\sigma}$ and $K_{p^{-1} \sigma}$ are the number of distinct loops making up each of the worldsheet boundary and the open string partition function assigns $N$ states for each loop, corresponding to the $N$ E-branes on which the open strings can end.    
Our final result is:
\begin{align}
\tr(\rho_V) =  \sum_{n} \frac{e^{\frac{-n \lambda A}{2}}}{n!} \sum_{i} \frac{(-1)^i}{i!}\left(\frac{\lambda A }{ N}\right)^{i} \sum_{\sigma \in S_{n}} \sum_{p_{1} \dots p_{i} }  N^{K_{\sigma}}N^{K_{p^{-1} \sigma}}.
\end{align}
This reproduces the expected expression \eqref{ZS2'} for $Z^+_{S^2}$ (up to a trivial relabelling $p \to p^{-1}$).
But we have now seen how it arises as a trace over the open string Hilbert space: it is a thermal partition function describing the stringy entanglement thermodynamics of the Hartle- Hawking state.  
In particular, the thermal entropy of this open string ensemble (treating $A$ as an inverse temperature) coincides with the entanglement entropy of the interval $V$.
Moreover, we see that the factors $N^{K_\sigma}$ and $N^{K_{p^{-1} \sigma}}$ have a statistical interpretation as counting the distinct Chan-Paton indices associated with the open strings.

\subsection{The zero area limit}
To see more explicitly how the sphere partition function counts open string states, it is instructive to consider the zero area limit of $Z^{+}_{S^2}$.\footnote{The zero area limit of the chiral Yang-Mills theory on any two-manifold gives a topological string theory that has been studied previously in ref.~\cite{Cordes:1994sd}.  }
Interpreted as a closed string amplitude, the zero area limit just gives the normalization $\braket{\Omega|\Omega}$.
But in the open string channel, the zero area limit calculates the dimension of the open string Hilbert space, which counts the number of open string edge modes:
\begin{align}\label{A0}
\lim_{A\ra 0} Z^{+}_{S^{2}} &=\tr_{V}(\bold{1}) =\sum_{n=1}^{\infty} \frac{1}{n!}  \sum_{\sigma \in S_{n}}  N^{2K_{\sigma}} =\sum_{n=1}^{\infty} \frac{1}{n!}  \sum_{k=1}^n \left[ n \atop k \right] N^{2k} = \sum_{n = 1}^\infty \binom{ N^2 + n - 1}{n}.
\end{align} 
Here $\left[ n \atop k \right]$ is the number of permutations in $S_n$ with $k$ cycles, also known as the (unsigned) Stirling number of the first kind.
In the last line we have used the identity that the generating function for $\left[ n \atop k \right]$ is given by the ``rising factorial''
\begin{equation}
x^{(n)} := \frac{(x + n - 1)!}{(x-1)!} = \sum_{k = 0}^n \left[ n \atop k \right] x^k.
\end{equation}

The formula \eqref{A0} appeared in \cite{Gross:1992tu}, where it was calculated in the Yang-Mills theory by means of orthogonal polynomials.
We now see that it has a natural interpretation in terms of the open string theory: it is the dimension of the open string Hilbert space, which counts the number of ways of assigning pairs of Chan-Paton factors $(i,j) = 1 \ldots N$ to $n$ open strings, accounting for the indistinguishability of open strings.
This is given by the number of \emph{weak compositions} of $n$ into at most $N^2$ parts:
\begin{align}
\dim \mathcal{H}_n &= [x^n] (1 - x)^{-N^2} = \binom{ N^2 + n - 1}{n},
\end{align}
where $[x^n]$ is the operator that extracts the coefficient of $x^n$.

Finally, note that expressing $\lim_{A\ra 0} Z^{+}_{S^{2}}  $ in the representation basis gives another formula for $ \dim \mathcal{H}_{n}$:
\bea \label{R}
\sum_{R \in Y_{n}} (\dim R)^2 = \dim \mathcal{H}_{n}
\eea
where $Y_{n}$ denotes the set of Young diagrams with $n$ boxes.
This formula can be understood as follows. 
The symmetric group $S_{n}$ acts on $\mathcal{H}_{n}$ by permuting the right (or left) endpoints of the open string states $\ket{IJ}$.
The corresponding irreducible representations of $S_{n}$ are obtained by symmetrizing/antisymmetrizing the Chan-Paton indices according to the diagram $R \in Y_{n}$, and a basis for each such representation is given by the matrix elements $R_{ab}$ of the representation:
\begin{equation}
\mathcal{H}_{n}^{R} = \bigoplus_{a,b} \ket{R,a,b}, \qquad \braket{U|R\,a,b} := R_{ab}(U).
\end{equation}
These representations are of dimension $(\dim R)^{2}$, so the left hand side of \eqref{R} merely counts the basis elements of $\mathcal{H}_{n}$ which block diagonalizes the action of $S_n$.  
This is the basis that diagonalizes the open string interaction Hamiltonian $H_{1}$, which also acts by permuting open string endpoints.

\subsection{The free chiral string}
\label{subsection:freestring}

Another approximation we can make to the chiral theory is to keep the area finite, but to neglect the interaction terms in the Hamiltonian \eqref{H'}.
This theory can be studied at the level of the path integral by simply restricting the sum over worldsheets to those without interaction branch points; this calculation was carried out in ref.~\cite{Taylor:1994zm}.
Generally when one truncates the path integral in some way there is no guarantee that the resulting expression continues to define a canonical partition function of the form $\tr e^{- \beta H}$. 
Here we show that the partition function without branch points does define a canonical partition function of noninteracting open strings.

Unfortunately, the resulting partition function does not constitute a useful approximation to the interacting chiral partition function. 
As pointed out in ref.~\cite{Gross:1992tu}, the partition function has non-negligible contributions from states with $n \sim N^2$ strings. 
For these states there is an enhancement of the interaction term coming from the large number of strings which competes with the explicit factor of $1/N$ and renders the interactions non-negligible.

In the free string theory, the Hamiltonian simply counts the number of open strings weighted by the Nambu-Goto term. 
Since there are $N^2$ different labels for the open string endpoints, this leads to a partition function:
\begin{equation} \label{freestring}
\log Z = -N^2 \log (1 - e^{\frac{-\lambda A}{2}}).
\end{equation} 
This is simply the partition function of $N^2$ harmonic oscillators, whose occupation numbers count the number of open strings with Chan-Paton factors $(i,j)$.
This result agrees with the sum over maps of worldsheets into the target space denoted $F_{O1}$ in ref.~\cite{Taylor:1994zm} .
The theory which reproduces this partition function includes only one chiral sector (\textit{i.e.} the worldsheets have only one orientation), and does not include interaction branch points.
The only singularities are the $\Omega$-points, whose role is to count the open strings.
In the next section we consider the generalization to a theory with two chiral sectors.

\section{The coupled theory}
\label{section:coupled}

So far we have focused on the stringy description of a single chiral sector of 2D Yang-Mills theory.
Gross and Taylor showed that the full Yang-Mills partition function on $S^2$ is described by a closed string theory containing world sheets of two distinct orientations.
In the sum over worldsheets one must account for a new type of singularity: \emph{orientation reversing tubes}.
These are additional singularities located only at the $\Omega$-points that connect string worldsheets of opposite orientation.
Each such tube connects a chiral and antichiral string of the same winding number around the $\Omega$-point, and comes with a factor of $-1/N^2$. 
Below we review the description of the closed string Hilbert space including both chiral sectors and show how the tube diagrams emerge from the left-right entanglement structure of the E-brane boundary state.
We will then give an open string interpretation of the tubes and show that they arise from a counting of open string states taking into account the unitarity constraint $U U^\dag = 1$.
Thus the orientation-reversing tube singularities arise as a natural feature of the open string kinematics.

\subsection{The coupled closed string Hilbert space}

As discussed briefly in section \ref{section:closedstrings}, our Hilbert space $\mathcal{H}^+$ captures only one chiral sector of the Yang-Mills theory on a circle.
In the representation basis, this sector contains representations whose Young diagrams have a fixed number of boxes as $N \to \infty$.
In the string picture, these correspond to states obtained by acting on the vacuum with string creation operators $a_k^\dag$ a finite number of times.
The full Hilbert space also contains states whose energies $C_{2} \sim N$ are of the same order, but are not captured in this description because they correspond to representations whose Young diagrams have a number of boxes that scales with $N$.
In particular, for any representation state $\ket{S} \in \mathcal{H}^+$ one can consider the conjugate representation state $\ket{\bar{S}} \in \mathcal{H}^{-}$.
Their closed string wave functions depend on  $U^\dagger$ and satisfy a conjugate Frobenius relation (\textit{c.f.} \eqref{Fr})
\begin{equation}\label{AF}
\braket{U|\bar S} = \sum_{\tau \in S_n} \frac{\chi_S(\tau)}{n!} \braket{U| \bar{\tau}},
\end{equation}
where the symmetrization rules for the Young diagrams are now applied to tensors of antifundamentals.
The number of boxes in the Young diagram for $S$ does not scale with $N$, so the expression \eqref{AF} has a well-defined limit as $N \ra \infty$.
The closed ``antistring'' states $\ket{\bar \tau}$ of this equation are defined as in  \eqref{CS}, but with traces of powers of $U^\dag$.
These states belong to $\mathcal{H}^{-}$ and can be visualized as collections of closed strings winding in the opposite direction around the spatial circle.

One can now construct a Hilbert space by combining the states $\ket{\sigma}$ and $\ket{\bar \tau}$.
However, this is not a simple tensor product: $U$ and $U^\dagger$ are not independent due to the unitarity condition $U U^\dag = 1$.
Treating the sectors as independent would therefore give an overcounting of the states.
For example, the state with zero strings (corresponding to the trivial representation) will appear in every tensor product $R \otimes \bar S$ which contains the trivial representation.

We can avoid this overcounting by summing over only \emph{coupled representations}.
The coupled representation $R \bar S$ is defined as the largest irreducible representation that appears in the tensor product of representations $R \otimes \bar S$.
Since these representations also have Young diagrams with $O(N)$ number of boxes, capturing their large $N$ limit requires a generalized version of the Frobenius relation. 
For $R$ and $S$ with $n$ and $\bar{n}$ boxes respectively, Gross and Taylor defined a set of coupled string states $\ket{\sigma, \bar \tau}$ such that
\begin{align}\label{GFR}
\ket{R\bar{S}}&= \sum_{\sigma,\tau} \frac{\chi_{R}(\sigma)\chi_{S}(\tau)}{n! \bar{n}!} \ket{\sigma ,\bar{\tau}}, \nn
\ket{\sigma ,\bar{\tau}}&= \sum_{R,S} \chi_{R}(\sigma) \chi_{S}(\tau)\ket{R\bar{S}},
\end{align}
where to leading order in $1/N$ the coupled closed string state $\ket{\sigma ,\bar{\tau}}$ has the wave function  
\bea\label{ccs}
\braket{U|\sigma ,\bar{\tau}} =\braket{U |\sigma}\braket{U|\bar{\tau}} + \cdots.
\eea
Substituting this leading approximation in  \eqref{GFR} gives $\ket{R \bar{S}} = \ket{R} \otimes \ket{\bar{S}}$ so at leading order the coupled representation $R\bar{S}$ can be treated as a tensor product of $R$ and $\bar{S}$.  
At this order the closed string Hilbert space factorizes into a product of strings and antistrings:
\bea
\lim_{N \rightarrow \infty} \mathcal{H}_{\U(N)} = \mathcal{H}^{+} \otimes \mathcal{H}^{-}.
\eea
The subleading corrections in \eqref{ccs} arise from subtracting the traces in the smaller irreducible representations that arise in $R \otimes \bar{S}$, creating entanglement between the two chiral sectors. 

The general form of these correction terms can be deduced from the Clebsch-Gordan rules for the Young diagrams, where the subleading terms are obtained from the leading one by ``annihilating'' strings and anti strings. 
The Clebsch-Gordan rules give the tensor product $R\otimes \bar{S}$ as a sum of representations whose diagrams are obtained from adding the boxes of $R$ to the boxes of $\bar{S}$.  
The largest of these representations is $R\bar{S}$, while the subleading ones correspond to different coupled representations $R' \bar{S}'$, in which $R'$ and $S'$ are obtained from $R$ and $S$ by deleting the same number $k$ of boxes from each.
These representations correspond to states which are products of $n-k$ and $\bar{n}-k$ strings and antistrings.
It is clear that the same coupled representation $R' \bar{S}'$ will occur in many different tensor products $R \otimes \bar{S}$, and the subleading terms in \eqref{ccs} correct for this overcounting.

However, a simpler way to derive these correction is to impose orthogonality of the coupled closed strings basis \cite{Gross:1993yt}:
\bea \label{oth}
\braket{\sigma' ,\bar{\tau}'|\sigma ,\bar{\tau} } = \delta_{T_{\sigma} T_{\sigma'}}\delta_{T_{\tau}T_{\tau'}} |C_{\sigma}| \, |C_{\tau}|,
\eea
where $T_\sigma$ denotes the orbit of the permutation $\sigma$, and the delta function ensures that $\sigma$ and $\sigma'$ belong to the same orbit i.e. they are conjugate, and $C_\sigma$ counts the number of permutations commuting with $\sigma$.
Equation \eqref{oth} is necessary for the consistency of \eqref{GFR} with the orthonormality of the characters and imposing it leads to an exact expression for the coupled closed string basis \cite{Gross:1993yt}:
\begin{align}\label{CCS}
\ket{\sigma, \bar{\tau}} &= \sum_{\nu} (-1)^{K_{\nu}} |C_{\nu}| \ket{\sigma \setminus \nu} \ket{\bar{\tau} \setminus \nu}, 
\end{align}
Above, $\nu$ is a set of cycles of $\sigma$ for which there is a corresponding set of cycles of $\tau$ of the same lengths.
The permutation $\sigma \setminus \nu$ is obtained by taking the cycles of $\sigma$ and deleting the set corresponding to $\nu$.   
$K_{\nu}$ is the number of cycles in $\nu$, which counts the number of strings and antistrings that have annihilated. 
Substituting \eqref{CCS} into \eqref{GFR} and choosing $U=U^{\dg}=1$ immediately leads to the dimension formula.
\bea \label{dimRSbar}
\dim R\bar{S} = \sum_{ \sigma, \tau}  \frac{\chi_{R}(\sigma) \chi_{S}(\tau)}{n!\bar{n}!}
\sum_\nu
(-1)^{K_{\nu}} |C_{\nu}| N^{K_{\sigma \setminus \nu} } N^{K_{\bar{\tau} \setminus \nu}}
\eea
Gross and Taylor used this formula to derive a large-$N$ expansion of the non-chiral partition function $Z_{S^2}$ and showed that that the subleading terms in $\frac{1}{N}$ such as \eqref{dimRSbar} could be expressed as string diagrams with orientation-reversing tubes.

\subsection{Entanglement tubes and the E-brane boundary state}

\begin{figure}
\centering
\includegraphics[scale=.25]{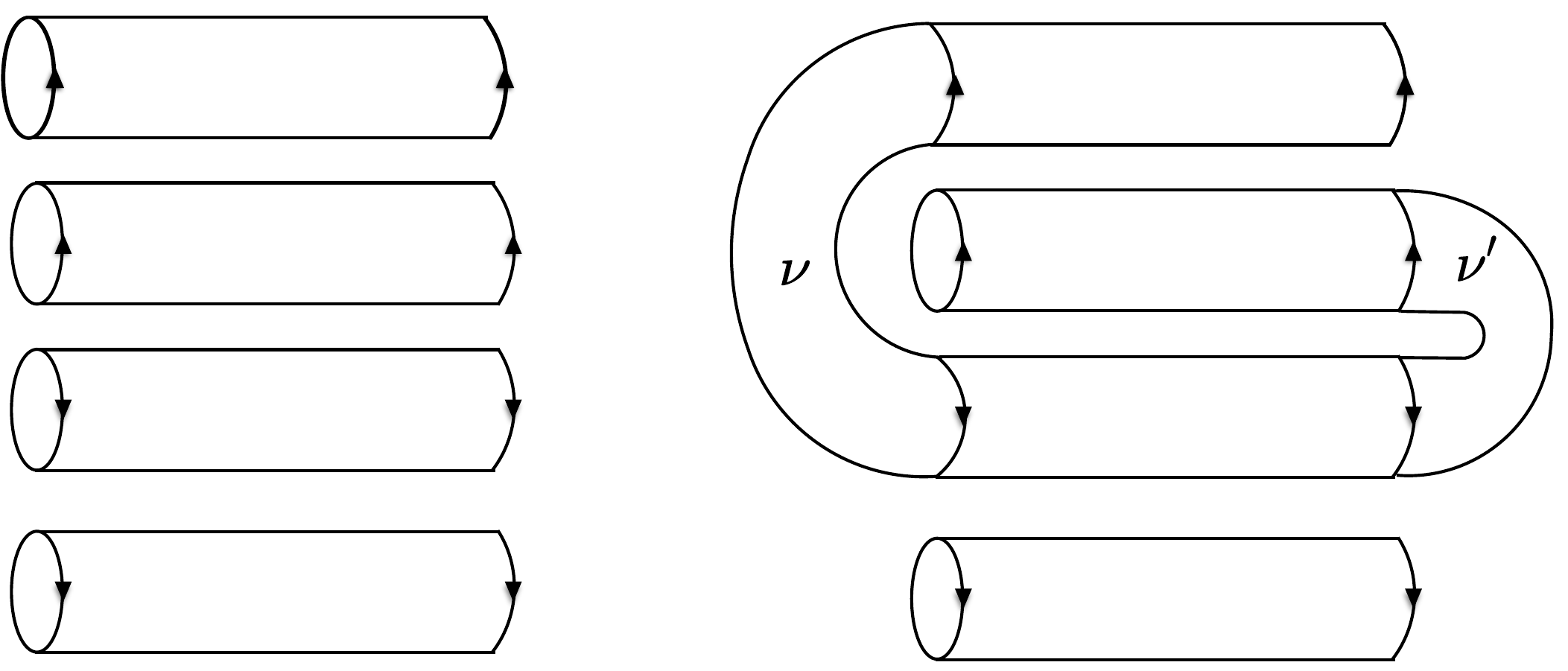}
\caption{Shown are two diagrams appearing in $ \lim_{A\ra 0} \bra{\Omega} e^{-\frac{\lambda A}{2 N} C_{2}} \ket{\Omega}$.  On the left,we have a decoupled closed string and anti string configuration.  In the physical Hilbert space, strings and anti strings  are entangled, leading to the diagram on the right where tubes $\nu$ and $\nu'$ connect string-anti string pairs. }
\label{tubes}
\end{figure}

Here we will show that the tubes connecting oppositely oriented strings in the diagram expansion for $Z_{S^2}$ arise due to the left-right entanglement in the E-brane boundary state $\ket{\Omega}$. 
 This is best illustrated in the zero area limit:
\begin{align}
\lim_{A \ra 0} Z_{S^{2}} &= \lim_{A \ra 0} \bra{\Omega} e^{-\frac{\lambda A}{2 N} C_{2}} \ket{\Omega} \nn
 \ket{\Omega} &= \sum_{R, \bar{S} } \dim R\bar{S} \ket{R\bar{S}}.
 \end{align} 
As before, we have defined the E-brane boundary state $\braket{U|\Omega}$ as the Euclidean path integral on an infinitesimal disk.  
In the leading large-$N$ approximation where $ \dim R\bar{S} = \dim R \dim S$, the boundary state factorizes:
\begin{align}
 \ket{\Omega} &\sim  \ket{\Omega^{+}} \ket{\Omega^{-}}, \nn
\braket{\sigma,\bar{\tau}|\Omega} &\sim \braket{\sigma|\Omega^{+}}\braket{\bar{\tau} |\Omega^{-}} = N^{K_{\sigma}}N^{K_{\tau}}.
\end{align}
The amplitude $\braket{\Omega|\Omega}$ also factorizes.
Expressing $\ket{\Omega^{\pm}}$ in the coupled basis \eqref{CCS} leads to a diagrammatic expansion for  $\braket{\Omega|\Omega}$ in which independent strings and antistrings propagate between $\Omega$-points living in the same sector. This is illustrated in the left figure in \eqref{tubes}.  
As before, the coupling of the closed string states to the $\Omega$-point in each sector gives a factor of $N=\frac{1}{g_{\text{string}}}$ per closed string.

These diagrams must be corrected to account for the fact that the E-brane boundary state entangles the chiral and anti chiral  sector.   Using the dimension formula, we find that 
\begin{align}
\ket{\Omega} &= \sum_{R, \bar{S} } \dim R\bar{S} \ket{R\bar{S}}\nn
&=\sum_{n,\bar{n}} \frac{1}{n! \bar{n}!} 
\sum_{\substack{\sigma \in S_{n} \\ \tau \in S_{\bar{n}}}}
\left( \sum_{\nu } (-1)^{K_\nu} |C_{\nu}| N^{K_{\sigma \backslash  \nu }}N^{K_{\tau \backslash  \nu }}\right) \ket{\sigma, \bar{\tau}},
\end{align} 
where $\nu$ is defined as before.  
The amplitude is corrected accordingly: 
\begin{align} \label{OO}
\lim_{A \ra 0} \bra{\Omega} e^{-\frac{\lambda A}{2 N} C_{2}} \ket{\Omega} 
&= \sum_{n,\bar{n}} \frac{1}{n! \bar{n}!}   
\sum_{\substack{\sigma \in S_{n}  \\ \tau \in S_{\bar{n}}}} 
\sum_{\nu,\nu'}
(-1)^{K_{\nu}}  (-1)^{K_{\nu' }} |C_{\nu}| |C_{\nu'}| N^{ K_{\sigma \setminus \nu} + K_{\tau \setminus \nu}  } N^{ K_{\sigma \setminus \nu'}+ K_{\tau \setminus \nu'}} 
\end{align}
For a fixed $\sigma \in S_{n} $, $ \tau \in S_{\bar{n}}$, the $\nu=0$ term describe the propagation of decoupled strings and anti -strings described by $\sigma $ and $\tau$, with the familiar coupling of $N=\frac{1}{g_{\text{string}}}$ per closed string (see left figure in \eqref{tubes}).
For $\nu \neq 0$, each cycle of $\nu$ is represented by a tube that connects an external string antistring pair at one of the $\Omega$-points, causing them to annihilate.\footnote{We can also think of the tube as part of a closed string worldsheet for a string that is emitted and absorbed by the same $\Omega$-point.} 
One such tube diagram is depicted in figure \ref{tubes}.
Each tube decreases the number of holes in the worldsheet by 2, leading to a string coupling of $N^{-2}=g_{\text{string}}^{2}$.
In addition, each of these annihilations comes with a factor of $(-1)$, and a 
factor of $|C_{\nu}|$ which reflects the way in which the tube changes the symmetry factor of the diagram. 

\subsection{Open string description of the entanglement tubes}

To find a canonical open string interpretation of the entanglement tubes, we begin by considering the large-$N$ Hilbert space $\mathcal{H}$ on the interval $V$.
In the representation basis, this Hilbert space is spanned by states $\ket{R \bar{S},a,b}$ whose wavefunctions are matrix elements in the coupled representations:
\begin{equation}
\braket{U | R \bar{S}, a, b} = [R\bar{S}]_{ab}(U),
\end{equation}
with $a,b = 1, \ldots, \dim(R \bar S)$ indices in the representation $R \bar S$.
We are implicitly invoking a large-$N$ version of the Peter-Weyl theorem, which at finite $N$ says that the Hilbert space is spanned by matrix elements of the irreducible representations.

\begin{figure}
\centering
\includegraphics[scale=.25]{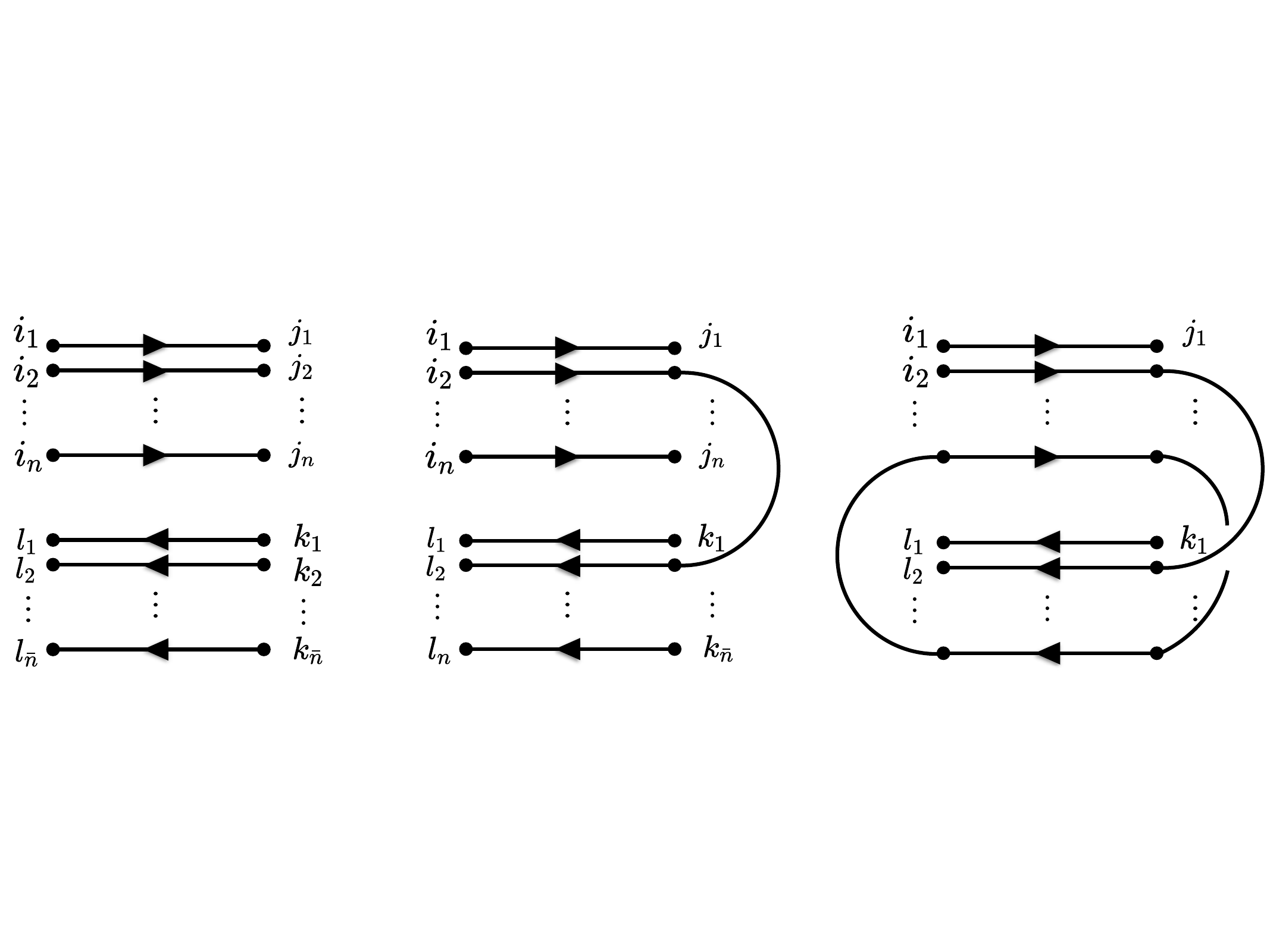}
\caption{On the left we have a diagram representing the states  $U_{IJ}U^{\dg}_{KL}$.  The diagram in the middle represents constraints imposed on $U_{IJ}U^{\dg}_{KL}$ obtained by contracting a pair of right endpoints.  The right most figure shows another diagram with contractions on both the left and right endpoints } 
\label{tubes1}
\end{figure}

In the leading order of the $1/N$ expansion where $R\bar{S} =R \otimes \bar{S}$, these matrix elements are obtained by symmetrizing/antisymmetrizing the open string states $\ket{IJ,KL}$ which we define as
\begin{equation}
\braket{U|IJ,KL} = U_{i_1 j_1} U_{i_2 j_2} \cdots U_{i_n j_n} U^\dag_{k_1 l_1} U^\dag_{k_2 l_2} \cdots U^\dag_{k_{\bar n} l_{\bar n}}.
\end{equation}
These can be represented, as shown in figure \ref{tubes1}, as a collection of $n$ open strings with Chan-Paton indices $(i_1, j_1), \ldots, (i_n, j_n)$ and $\bar n$ open antistrings with Chan-Paton indices $(k_1, l_1), \ldots, (k_{\bar n}, l_{\bar n})$.
The mixed symmetrization of Chan-Paton indices implements a projection onto the irreducible representations $R$ and $\bar S$ in each chiral sector, so summing over $R$ and $S$ removes this projection and results in a leading order Hilbert space $\mathcal{H}^{0}$:
\begin{align} \label{H0}
\mathcal{H}^{0} &=\bigoplus_{R,S,a,b } \ket{R \otimes \bar{S},a,b} = \bigoplus_{IJ,KL} \ket{IJ,KL}
\end{align} 
where there are no symmetries imposed on the Chan-Paton indices. 
$\mathcal{H}^{0}$ is an extended Hilbert space in which $U$ and $U^\dag$ are formally treated as independent matrices not constrained by the identity $U U^{\dagger} = 1$.  
The subspace of physical states $\mathcal{H} \subset  \mathcal{H}^{0}$ is obtained by projecting onto the irreducible representation $R\bar{S} \subset R \otimes \bar{S}$ in each term of the sum in \eqref{H0}.  
This projection is equivalent to enforcing the constraint  $U U^{\dagger}=1$ on  $\mathcal{H}^{0}$, thereby eliminating the linear dependence between the states $\ket{IJ,KL}$ with different numbers of open strings.

To see why projection onto coupled representations is equivalent to the unitarity constraint, consider the space of matrix elements of $R \otimes \bar S$, where $R$ is an irreducible representation whose Young diagram has $n$ boxes, and $S$ is an irreducible representation whose Young diagram has $\bar n$ boxes.
These states transform as $\U(N)$ tensors, so we may decompose them into irreducible representations of $\U(N)$ by standard methods.
This simply amounts to subtracting out all possible traces which contract a fundamental index with an antifundamental index.

Let $\bold{P}$ denote the projection operator that projects each space $R \otimes \bar{S}$ to the coupled representation $R \bar S$.
We can now see that the zero area limit of $Z_{S^2}$ is simply expressing a trace over the open string Hilbert space:
\begin{align}\label{ZO}
\lim_{A\ra 0} Z_{S^{2}} &=\tr_{\mathcal{H}}(\bold{1})\nn
&=\tr_{\mathcal{H} ^{0} }(\bold{P})\nn
 &=
 \sum_{n,\bar{n}} \frac{1}{n! \bar{n}!}  
 \sum_{\substack{\sigma \in S_{n} \\ \tau \in S_{\bar{n}}}}
 \sum_{\nu,\nu'} (-1)^{K_{\nu}}  (-1)^{K_{\nu' }} |C_{\nu}| |C_{\nu'}| N^{ K_{\sigma \setminus \nu} + K_{\tau \setminus \nu}  } N^{ K_{\sigma \setminus \nu'}+ K_{\tau \setminus \nu'}}.
\end{align}
This dimension formula can be understood by implementing the projection $\bold{P}$ systematically on the overcomplete set of states $\ket{IJ,KL}$ in each sector of fixed $n$ and $\bar{n}$.  

\begin{figure}
\centering
\includegraphics[scale=.25]{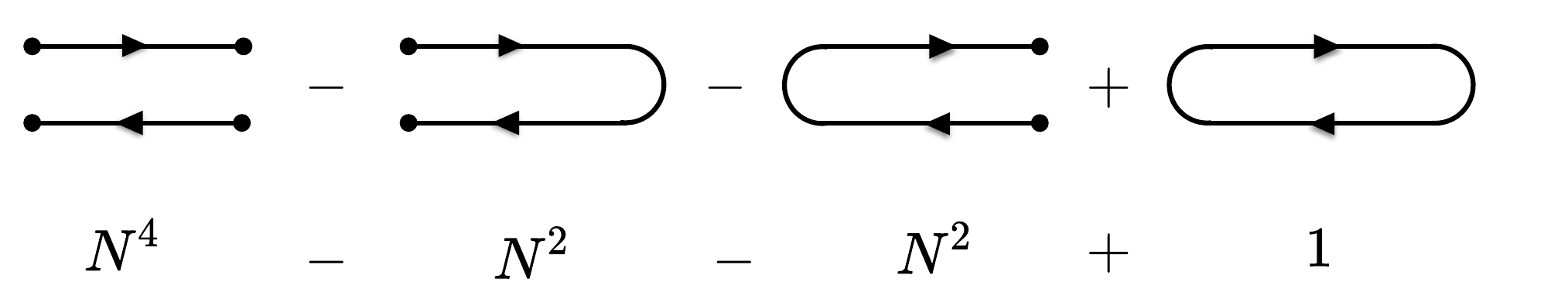}
\caption{The left figure illustrates the diagrammatic computation for the number of  open string edge modes for $n=\bar{n}=1$.
As expected, this gives the square of the dimension of the adjoint representation, $(N^2 - 1)^2$.}
\label{Adjoint}
\end{figure}

The formula \eqref{ZO} is best illustrated by way of examples: let us first consider the case $n=\bar{n} =1 $.
To apply $\bold{P}$ to the state $\ket{ij,kl}$, we first subtract the tensor obtained from  $U_{ij}U^{\dg}_{kl} $ by contracting one pair of indices, 
\bea
U_{ij} U^\dag_{kl} \to U_{ij}U^{\dg}_{kl} - \frac{\delta_{jk}}{N}  U_{ia}U^{\dg}_{al}.
\eea
We then do the same for the remaining two indices:
\bea\label{states}
\braket{U|\bold{P} | ij, kl }= \left( U_{ij}U^{\dg}_{kl} - \frac{\delta_{jk}}{N}  U_{ia}U^{\dg}_{al}\right) - \frac{\delta_{il}}{N}  U_{aj}U^{\dg}_{ka}  + \frac{\delta_{jk} \delta_{il}}{N^2}    U_{ba}U^{\dg}_{ab}.
\eea
To see how this leads to the dimension formula \eqref{ZO}, we need to count the number of independent states in \eqref{states}, which amounts to enumerating the number of independent constraints we have imposed on the $N^4$ initial states due to the condition $U^{\dag}U=1 $. This can be done iteratively as follows.  Start with the constraints involving one contraction: 
\bea\label{1} 
U_{ia}U^{\dg}_{al}=\delta_{i l},  \\
U_{aj}U^{\dg}_{ka}=\delta_{j k}. \label{2}
\eea  
This gives a total of $2 N^2$ constraints. 
However these constraints are not independent because the constraint corresponding to contracting both indices simultaneously,
\bea\label{3}
U_{ab}U^{\dg}_{ba}=N,
\eea
is counted in both  \eqref{1} and \eqref{2}.
Thus the total number of independent states is 
\bea
\dim \mathcal{H}_{1,1} = N^{4}- N^{2}-N^{2} +1 .
\eea 
We can associate a diagram to each set of constraints in \eqref{1}-\eqref{3} with an open string for each $U$ and oppositely oriented antistring for each $U^\dg$. 
Each contraction is represented by a line connecting a pair of open string endpoints as in figure  \eqref{Adjoint}. 
Since these lines correspond to slices through the tube diagram in the closed string picture, we will also refer to them as tubes.  
Each tube reduces the number of free endpoints by $2$, and so decreases the number of states in the diagram by $N^2$.
It also carries a factor of $-1$ arising from the fact that we are \emph{subtracting} the trace.
This counting is illustrated in figure \ref{Adjoint}, and indeed yields the correct number of states with $n = \bar n = 1$.

When $n$ or $\bar{n}$ is greater than $1$, associating $N$ states per Chan-Paton index will overcount the number of states because of open string indistinguishability. 
More precisely, the string-antistring states are invariant under the action of the permutation group $S_{n} \times S_{\bar{n}}$ which relabels the strings within the same sector.
Thus in performing the trace in \eqref{ZO} we should sum over all unrestricted Chan-Paton indices and divide by their orbit under this symmerty group.
As in the chiral case, this will lead to a sum over diagrams in which Chan-Paton indices are correlated, and an overall symmetry factor of $n! \bar{n}! $.

For example, the number of independent basis states contained $\ket{IJ,KL}$ prior to the projection is: 
 \begin{align}
\dim \mathcal{H}_{n,\bar{n}}^{0}
&= \sum_{I,J,K,L}  \frac{1}{|T_{IJ,KL}|} 
\nn
&= \sum_{I,J,K,L}   \sum_{\sigma \in S_n}  \sum_{\tau \in S_{\bar{n}}}  \frac{\delta(I, \sigma(I)) \, \delta (J, \sigma (J)) \, \delta(K, \tau(K)) \, \delta (L, \tau (L))}{|T_{IJ,KL}| |C_{IJ,KL}|}    \\
&= \frac{1}{n!}  \frac{1}{\bar{n}!}  \sum_{\sigma \in S_n}  \sum_{\tau\in S_{\bar{n}} }N^{ 2 K_\sigma}  N^{2 K_\tau} .
\end{align}
Here $T_{IJ,KL}$ is the orbit of the open string state under $S_n \times S_{\bar n}$, and $C_{IJ,KL}$ its stabilizer.
This reproduces the leading term in \eqref{ZO}, in which $\nu = \nu' = \emptyset$.

The subleading terms in \eqref{ZO} accounts for indistinguishability at subsequent steps in the iterative counting of the constraints imposed by $\bold{P}$. 
For example, consider the constraints where a right index of an open string is contracted with a left index of an antistring:
\begin{align}\label{T1}
\sum_{a} U_{i_{1} a} \cdots U_{ i_{n} j_{n} }   U^{\dg}_{ a l_{1} } \cdots  U^{\dg}_{k_{\bar{n}}  l_{\bar{n}}}  =\delta_{i_{1}l_{1}}U_{i_{2} j_{2}} \cdots U_{ i_{n} j_{n} }   U^{\dg}_{ k_{2} l_{2} } \cdots  U^{\dg}_{k_{\bar{n}}  l_{\bar{n}}}.
\end{align}
Just like the string states, the constraints can be labelled by multi-indices $(I,J)$ and $(K,L)$, except that $J$ and $K$ are allowed to contain one contracted index $a$.  These indices are again invariant under the action of the symmetric group $S_{n} \times S_{\bar{n}}$, so we can write the number of  non-identical constraints in \eqref{T1} as 
\begin{align}
\sum_{I,J,K,L}  \frac{1}{|T_{IJ,KL}|} 
&= \sum_{I,J,K,L}   \sum_{\sigma \in S_n}  \sum_{\tau \in S_{\bar{n}}}  \frac{\delta(I, \sigma(I)) \, \delta (J, \sigma (J)) \, \delta(K, \tau(K)) \, \delta (L, \tau (L))}{|T_{IJ,KL}| |C_{IJ,KL}|} \\
&= \frac{1}{n!\bar{n}!} \sum_{\sigma \in S_n}  \sum_{\tau \in S_{\bar{n}}} | \text{Fix} (\sigma,\tau )   |.
 \end{align}
The fix point set $ \text{Fix} (\sigma,\tau )$ consists of equations of the general form \eqref{T1} that are invariant under $\sigma \times \tau$. 

These are equations where the contracted index $a$ belongs to a 1-cycle in $\sigma$ and $\tau$ respectively.  Meanwhile non-contracted indices must take the same value on each cycle of $\sigma$ and $\tau$, so there are  $N^{2K_{\sigma}-1}N^{2K_{\tau}-1}$ elements in  $ \text{Fix} (\sigma,\tau )$.  This counting gives the terms in \eqref{ZO} in which  $\nu$ is a cycle of length one and corresponds to all diagrams in which a single tube appears on the right side of the open strings.  These diagrams are depicted in the middle in figure  \ref{tubes1} .

At the next order in $1/N$ we must account for the fact that the constraints themselves are not independent.
The equations \eqref{T1} are redundant because they satisfy relations obtained by contracting additional pairs of indices. 
For example, for $n=\bar{n}=2$ the $N^{6}$ constraints 
\bea \label{N6}
\sum_{a} U_{i_{1} a}U_{ i_{2} j_{2} }   U^{\dg}_{ a l_{1} } U^{\dg}_{k_{2}  l_{2}}  =\delta_{i_{1}l_{1}}U_{i_{2} j_{2}}    U^{\dg}_{ k_{2} l_{2} } 
\eea
are related by the $\frac{1}{2}(N^{4} + N^2)$ equations
\bea \label{T2}
  U_{i_{1} a }   U^{\dg}_{ a l_{1} }  U_{i_{2} b}   U^{\dg}_{ b l_{2} } =\delta_{i_{1}l_{1}} \delta_{i_{2}l_{2}}.
\eea
However, to count the number of independent relations in \eqref{T2}, we must take care not to include the equations with $i_{1}=i_{2}$ and $l_{1}=l_{2}$, which take the form
\bea\label{sq}
U_{i a}U^{\dg}_{al }  U_{i b}U^{\dg}_{bl } = \delta_{il} \delta_{il}.
\eea
This is just the square of a constraint on $\mathcal{H}^{0}_{1,1}$, so it should not be counted in $\mathcal{H}^{0}_{2,2}$.  Excluding these $N^2$ equations, we conclude that the number of independent constraints in \eqref{N6} on the subspace $\mathcal{H}^{0}_{2,2}$ is  
\bea\label{in}
 N^{6} - (\frac{1}{2}(N^{4} + N^2) - N^2).
\eea
These are represented by the $n=\bar{n}=2$ terms in \eqref{ZO} with nontrivial ``tube"  $\nu$ : 
\bea
\frac{1}{2! \, 2!} \sum_{(\sigma, \tau) \in S_{2}\times S_{2}  } \sum_{\nu \neq \emptyset} (-1)^{K_{\nu}} |C_{\nu}| N^{K_{\sigma}}N^{K_{\tau}} N^{ K_{\sigma \setminus \nu} + K_{\tau \setminus \nu}  }  = - \frac{1}{4} ( 4N^{6}  - 2N^4 +  2 N^2)
\eea
where the $N^6$ leading term comes from $\nu$ a single 1-cycle, the $N^{4}$ correction corresponds to terms with $\nu$ a pair of 1-cycles and the $N^2$ term corresponds to $\nu$ being a 2-cycle.  
Comparing with \eqref{in}  we see that the symmetry factor $C_{\nu}=2$  and the minus sign for the $\nu=(2)$ is exactly what's needed to exclude the $N^2$  redundant constraints in \eqref{sq}.

Including also the contractions of the other index, corresponding to terms with nontrivial $\nu'$, gives the dimension of the space $\mathcal{H}_{2,2}$.
Thus the alternating sign structure of \eqref{ZO} arises from the iterative counting of constraints imposed by the projection operator $\bold{P}$, and the zero area limit of the sphere partition enumerates the number of open string edge modes.  
 
\subsection{Entanglement and modular Hamiltonian}

Having described the Hilbert space of the non-chiral string theory, all that remains to understand the entanglement entropy is the modular Hamiltonian.
This is given by a multiple of the quadratic Casimir:
\bea
H_{V}= \frac{\lambda A}{2 N} C_{2} 
\eea
where $C_{2}$ is defined as in \eqref{C2}. 
To see how this operator acts on the open string states, it is useful to lift it to the space $\mathcal{H}^0$ on which $U$ and $U^\dag$ are independent.
Applying the relation \eqref{partial}, we find that the Casimir can be written as a sum of chiral and antichiral pieces, plus a coupling term between the two sectors:
\begin{align} \label{Cc}
C_{2} &= C_{+} + C_{-}  + C_{c}, \nn
C_{+} &= N U_{ik} \frac{\partial}{\partial U_{ik}} + U_{ik} U_{jl} \frac{\partial}{\partial U_{jk}} \frac{\partial}{\partial U_{il}},
 \nn
C_{c} &= - 2 \left( U_{il} \frac{\partial}{\partial U_{ik}}  \right)\left( U^{\dagger}_{lj} \frac{\partial}{\partial U^{\dagger}_{kj}}\right).
\end{align} 
$C_{-}$ has the same functional form as the $C_{+}$, but with $U \to U^\dag$.
Note that these operators are acting on $\mathcal{H}^0$, where the derivatives $ \frac{\partial} {\partial U_{ij}}$ and  $ \frac{\partial}{\partial U^{\dg}_{ij}}$ act independently.
Here we see that the modular Hamiltonian contains all the terms expected from the worldsheet expansion.
The leading term in $C_\pm$ leads to the Nambu-Goto area term, and the subleading term implements the branch point singularities that couple two open strings of the same chirality.

The coupling term $C_c$ generates an interaction between chiral and antichiral strings, and its presence ensures that $C_2$ commutes with the projector $\bold{P}$. One might wonder why there are no diagrams associated with the term $C_c$, which would correspond to local interactions coupling worldsheets of opposite orientations. 
The reason these diagrams do not appear is that the string states created by $C_c$ contain a trace $U_{il} U^\dag_{lj}$, and hence are annihilated by the projector $\bold{P}$.

The sphere partition function can be expressed as
\bea
Z_{S^2} = \tr_{\mathcal{H}^{0} }(\bold{P}e^{- \frac{\lambda A}{2 N} C_{2} } ).
\eea
This completes our statistical accounting for the entanglement entropy of the Hartle-Hawking state in the Gross-Taylor model.
The entanglement entropy is given by the thermal entropy of a gas of non-chiral open strings with endpoints anchored on two stacks of E-branes located at the entangling surface, with interactions given by \eqref{Cc}.

\subsection{The non-chiral free string}

As we did in section \ref{subsection:freestring} for the chiral theory, we can consider the truncation of the coupled theory where interactions are neglected.
This corresponds to a path integral over string worldsheets with two distinct orientations on a sphere with two $\Omega$-point singularities, and orientation-reversing tubes.
This path integral was carried out in ref.~\cite{Taylor:1994zm}, and the result is
\begin{equation}
\log Z = 2 N^2 \sum_{m = 1}^\infty (-1)^m \log (1 - e^{- m A/2}) = 2 N^2 \log  \left( \sum_{n = 0}^{\infty} e^{-n (n+1) A / 4} \right).
\end{equation}
As in the chiral case, this partition function does correspond to a canonical sum over states: it is the partition function of $2 N^2$ uncoupled copies of a theory whose energy levels are the triangular numbers $T_n = n (n+1)/2$.

However, unlike the case of the chiral theory we do not have a description of the energy eigenstates that appear in this partition function.
Unlike the case of the chiral theory, they are not simply states with a definite number of open strings.
However we note that the partition function can be expressed in terms of a Jacobi $\vartheta$ function as 
\begin{equation}
\log Z = 2 N^2 \log \left(\frac{\vartheta_2(0,e^{-A/4})}{2 e^{-A/16}} \right),
\end{equation}
suggesting a possible connection to modular invariance.
We leave the canonical description of these states as an open puzzle.

\section{Discussion and Future work}
\label{section:discussion}

We have shown how a description of entanglement between regions of space in the Gross-Taylor model necessitates a formulation of the theory in terms of open strings.
We have shown that the entanglement entropy in the Gross-Taylor model counts open strings: more precisely the $\Omega$-point singularities discovered in ref.~\cite{Gross:1993yt} count the number of distinct open string states accounting for both the indistinguishability of open strings, and the unitarity constraint.
In the process, we have uncovered the E-brane; an object which sits at the entangling surface on which open strings can end.

Perhaps the most interesting product of our analysis is the relation between branes and entanglement.
In a sense the effect of the E-brane is essentially just to change the statistical weight of certain configurations. 
However, the E-brane we have found acts in many respects like a D-brane, for example both are nonperturbative objects. 
It was shown in ref.~\cite{Polchinski:1994fq} that the partition function in the presence of a D-brane state is of order $e^{-1/g_\text{string}}$ and hence D-branes are nonperturbative objects.
The $E$-branes have this same property: the partition function on the sphere satisfies $\log Z = O(N^2)$, and hence $Z \sim e^{-N^2} = e^{-1/g_\text{string}^2}$ where we have identified $g_\text{string} = 1/N$. 
The square of the coupling reflects the fact that the sphere partition function has two E-branes.
Hence E-branes are nonperturbative objects in the same sense that D-branes are.

An important question is to understand the dynamics of E-branes from the worldsheet perspective.
One interesting clue from the sum over worldsheets is that the position of the E-branes is not integrated over.
This is reminiscent of a D-brane, which is associated with Dirichlet boundary conditions that fix the location of the open string endpoints in spacetime.
However our two-dimensional model is too simple to answer any more detailed dynamical questions.
Due to the area-preserving diffeomorphism symmetry of the theory, the precise location of the E-brane is not a gauge-invariant concept.
Moreover, the E-brane in this case has no transverse dimensions, leaving unanswered the question of how the transverse coordinates of open strings should be treated for higher-dimensional E-branes.
Addressing these dynamical questions would necessitate studying a higher-dimensional string theory with local degrees of freedom.

There remain open questions even about this simple two-dimensional model.
One such question is the description of entangling surfaces consisting of multiple intervals.
The closed string formulation of two-dimensional Yang-Mills theory holds on manifolds of higher genus as well, except that one must introduce $2g - 2$ ``$\Omega^{-1}$-points''.
These $\Omega^{-1}$-points are analogous to the $\Omega$-points in that they allow for arbitrary singularities, but they are weighted differently in the path integral.
We will not consider the higher genus case here, but it arises naturally in the calculation of entanglement entropy for multiple intervals.
In that case, we do not expect to have a simple geometric description of the modular Hamiltonian as we do on the sphere, since the higher genus surfaces do not admit a foliation by intervals.
Nevertheless, we expect to be able to describe the modular Hamiltonian as a combination of a geometric evolution, together with the insertion of topology-changing operators corresponding to the $\Omega^{-1}$-points.

An interesting phenomenon that appears on the sphere is the Douglas-Kazakov phase transition \cite{Douglas:1993iia}.
On the sphere there is a competition between the dimension term in the partition function which favors states with a large number of strings, and the quadratic Casimir  which favors states with a small number of strings.
Above the critical temperature, the string states no longer provide a good description.
Given that we have identified that suitable Hilbert space of states in which the sphere partition function defines a canonical partition function, it would be of interest to find the states that provide a suitable weakly coupled description beyond the Douglas-Kazakov phase transition.

Another remaining open question is to find a $\sigma$-model Lagrangian whose partition function yields the sum over maps (see e.g. the discussion after eq (2.4) of ref.~\cite{Gross:1992tu}).
Ho\v{r}ava proposed such a string sigma model description in ref.~\cite{Horava:1993aq}.  
It would be interesting to see if these $\sigma$ models could be generalized to open strings with endpoints anchored at the $\Omega$-points. 
This would help us to understand spacetime entanglement from the perspective of the string worldsheet, which will be important in treating higher dimensional string theories which don't have a simple string field theory description as in the case of the Gross-Taylor string.  

A key question is whether this two-dimensional toy model can be used to gain insight into string theory entanglement in models with more spacetime dimensions.
One possible way forward is via gauge-gravity duality: in the Hamiltonian formulation, lattice Yang-Mills theory is simply many copies of two-dimensional Yang-Mills theory, each defined on the links of the lattice, with couplings at the lattice sites.
Thus at large $N$ we can describe lattice Yang-Mills theory as a theory of strings.
This approach was pursued in ref.~\cite{Lee:2010ub} to understand the emergence of a dual bulk.
The open string description presented here may be useful in understanding the proposed duality for subregions \cite{Dong:2016eik}.

Ultimately we would like to understand entanglement in higher-dimensional string theories with local degrees of freedom.
One existing approach to this question is via gauge-gravity duality.
In ref.~\cite{Faulkner:2013ana} it was shown that subleading corrections to the Ryu-Takayanagi formula calculate bulk entanglement entropy via the replica trick.
This derivation makes use of a brane-like surface extending into the bulk that acts as an entangling surface for the bulk field theory; it would be interesting to understand the coupling of strings to this surface, and whether it can be understood as counting states of the string endpoints.

We have shown that the Gross-Taylor model provides a precise realization of Susskind's picture of entanglement entropy in string theory arising from genus-0 closed string diagrams.
Here we see clearly that the leading order $N^2$ scaling of the entropy, which is necessary to obtain agreement with the $\sim 1/G$ scaling of the Bekenstein-Hawking entropy, comes from the statistical weight of the Chan-Paton factors associated with open strings.
This would seem to support the picture, suggested in refs.~\cite{Donnelly:2016auv,Harlow:2016vwg} that the Bekenstein-Hawking term arises from bulk entanglement  entropy, accounting for the appropriate (in this case stringy) edge modes.

\section*{Acknowledgments}

We thank Aron Wall for his comments, and for coining the term E-brane; David Gross and Wadi Taylor for discussions on the string formulation of Yang-Mills theory; Ben Michel for discussions on D-branes; and finally Joe Polchinski for many helpful discussions and for inviting GW to KITP.
WD is supported by the Department of Energy under Contract DE-SC0011702.
GW is supported by KITP graduate fellowship and the Presidential Fellowship from University of Virginia.

\bibliographystyle{utphys}
\bibliography{2dym}

\providecommand{\href}[2]{#2}\begingroup\raggedright\begin{thebibliography}{10}

\bibitem{Sorkin1983}
R.~D. Sorkin, ``On the entropy of the vacuum outside a horizon,'' in {\em Tenth
  International Conference on General Relativity and Gravitation (held Padova,
  4-9 July, 1983), Contributed Papers}, vol.~2, pp.~734--736.
\newblock 1983.
\newblock \href{http://arxiv.org/abs/1402.3589}{{\ttfamily arXiv:1402.3589
  [gr-qc]}}.

\bibitem{Bombelli1986}
L.~Bombelli, R.~K. Koul, J.~Lee, and R.~D. Sorkin, ``Quantum source of entropy
  for black holes,'' \href{http://dx.doi.org/10.1103/PhysRevD.34.373}{{\em
  Phys. Rev. D} {\bfseries 34} no.~2, (1986) 373--383}.

\bibitem{Srednicki1993}
M.~Srednicki, ``Entropy and area,''
  \href{http://dx.doi.org/10.1103/PhysRevLett.71.666}{{\em Phys. Rev. Lett.}
  {\bfseries 71} (1993) 666--669},
\href{http://arxiv.org/abs/hep-th/9303048}{{\ttfamily arXiv:hep-th/9303048}}.

\bibitem{Susskind:1993ws}
L.~Susskind, ``Some speculations about black hole entropy in string theory,''
\href{http://arxiv.org/abs/hep-th/9309145}{{\ttfamily arXiv:hep-th/9309145
  [hep-th]}}.

\bibitem{Susskind:1994sm}
L.~Susskind and J.~Uglum, ``Black hole entropy in canonical quantum gravity and
  superstring theory,'' \href{http://dx.doi.org/10.1103/PhysRevD.50.2700}{{\em
  Phys. Rev.} {\bfseries D50} (1994) 2700--2711},
\href{http://arxiv.org/abs/hep-th/9401070}{{\ttfamily arXiv:hep-th/9401070
  [hep-th]}}.

\bibitem{Dabholkar:1994ai}
A.~Dabholkar, ``Strings on a cone and black hole entropy,''
  \href{http://dx.doi.org/10.1016/0550-3213(95)00050-3}{{\em Nucl. Phys.}
  {\bfseries B439} (1995) 650--664},
\href{http://arxiv.org/abs/hep-th/9408098}{{\ttfamily arXiv:hep-th/9408098
  [hep-th]}}.

\bibitem{Dabholkar:1994gg}
A.~Dabholkar, ``Quantum corrections to black hole entropy in string theory,''
  \href{http://dx.doi.org/10.1016/0370-2693(95)00056-Q}{{\em Phys. Lett.}
  {\bfseries B347} (1995) 222--229},
\href{http://arxiv.org/abs/hep-th/9409158}{{\ttfamily arXiv:hep-th/9409158
  [hep-th]}}.

\bibitem{Larsen1995}
F.~Larsen and F.~Wilczek, ``Renormalization of black hole entropy and of the
  gravitational coupling constant,''
  \href{http://dx.doi.org/10.1016/0550-3213(95)00548-X}{{\em Nucl.Phys.}
  {\bfseries B458} (1996) 249--266},
\href{http://arxiv.org/abs/hep-th/9506066}{{\ttfamily arXiv:hep-th/9506066
  [hep-th]}}.

\bibitem{Solodukhin1995}
S.~N. Solodukhin, ``One loop renormalization of black hole entropy due to
  nonminimally coupled matter,''
  \href{http://dx.doi.org/10.1103/PhysRevD.52.7046}{{\em Phys. Rev. D}
  {\bfseries 52} (1995) 7046--7052},
  \href{http://arxiv.org/abs/hep-th/9504022}{{\ttfamily arXiv:hep-th/9504022
  [hep-th]}}.

\bibitem{Kabat:1995eq}
D.~N. Kabat, ``Black hole entropy and entropy of entanglement,''
  \href{http://dx.doi.org/10.1016/0550-3213(95)00443-V}{{\em Nucl. Phys.}
  {\bfseries B453} (1995) 281--299},
\href{http://arxiv.org/abs/hep-th/9503016}{{\ttfamily arXiv:hep-th/9503016
  [hep-th]}}.

\bibitem{He:2014gva}
S.~He, T.~Numasawa, T.~Takayanagi, and K.~Watanabe, ``Notes on entanglement
  entropy in string theory,''
  \href{http://dx.doi.org/10.1007/JHEP05(2015)106}{{\em JHEP} {\bfseries 05}
  (2015) 106},
\href{http://arxiv.org/abs/1412.5606}{{\ttfamily arXiv:1412.5606 [hep-th]}}.

\bibitem{Fursaev1996}
D.~V. Fursaev and G.~Miele, ``Cones, spins and heat kernels,''
  \href{http://dx.doi.org/10.1016/S0550-3213(96)00631-1}{{\em Nucl.Phys.}
  {\bfseries B484} (1997) 697--723},
\href{http://arxiv.org/abs/hep-th/9605153}{{\ttfamily arXiv:hep-th/9605153
  [hep-th]}}.

\bibitem{Donnelly2011}
W.~Donnelly, ``Decomposition of entanglement entropy in lattice gauge theory,''
  \href{http://dx.doi.org/10.1103/PhysRevD.85.085004}{{\em Phys.Rev.}
  {\bfseries D85} (2012) 085004},
\href{http://arxiv.org/abs/1109.0036}{{\ttfamily arXiv:1109.0036 [hep-th]}}.

\bibitem{Donnelly:2014fua}
W.~Donnelly and A.~C. Wall, ``Entanglement entropy of electromagnetic edge
  modes,'' \href{http://dx.doi.org/10.1103/PhysRevLett.114.111603}{{\em Phys.
  Rev. Lett.} {\bfseries 114} no.~11, (2015) 111603},
\href{http://arxiv.org/abs/1412.1895}{{\ttfamily arXiv:1412.1895 [hep-th]}}.

\bibitem{Donnelly:2015hxa}
W.~Donnelly and A.~C. Wall, ``Geometric entropy and edge modes of the
  electromagnetic field,''
\href{http://arxiv.org/abs/1506.05792}{{\ttfamily arXiv:1506.05792 [hep-th]}}.

\bibitem{Donnelly2012}
W.~Donnelly and A.~C. Wall, ``Do gauge fields really contribute negatively to
  black hole entropy?,''
  \href{http://dx.doi.org/10.1103/PhysRevD.86.064042}{{\em Phys.Rev.}
  {\bfseries D86} (2012) 064042},
\href{http://arxiv.org/abs/1206.5831}{{\ttfamily arXiv:1206.5831 [hep-th]}}.

\bibitem{Donnelly2014a}
W.~Donnelly, ``Entanglement entropy and nonabelian gauge symmetry,''
  \href{http://dx.doi.org/10.1088/0264-9381/31/21/214003}{{\em Class. Quantum
  Grav.} {\bfseries 31} no.~21, (2014) 214003},
  \href{http://arxiv.org/abs/1406.7304}{{\ttfamily arXiv:1406.7304 [hep-th]}}.

\bibitem{Cordes:1994fc}
S.~Cordes, G.~W. Moore, and S.~Ramgoolam, ``{Lectures on 2D Yang-Mills theory,
  equivariant cohomology and topological field theories},''
  \href{http://dx.doi.org/10.1016/0920-5632(95)00434-B}{{\em Nucl. Phys. Proc.
  Suppl.} {\bfseries 41} (1995) 184--244},
\href{http://arxiv.org/abs/hep-th/9411210}{{\ttfamily arXiv:hep-th/9411210
  [hep-th]}}.

\bibitem{Donnelly:2014gva}
W.~Donnelly, ``Entanglement entropy and nonabelian gauge symmetry,''
  \href{http://dx.doi.org/10.1088/0264-9381/31/21/214003}{{\em Class. Quant.
  Grav.} {\bfseries 31} no.~21, (2014) 214003},
\href{http://arxiv.org/abs/1406.7304}{{\ttfamily arXiv:1406.7304 [hep-th]}}.

\bibitem{Gromov:2014kia}
A.~Gromov and R.~A. Santos, ``Entanglement entropy in {2D} non-abelian pure
  gauge theory,'' \href{http://dx.doi.org/10.1016/j.physletb.2014.08.023}{{\em
  Phys. Lett.} {\bfseries B737} (2014) 60--64},
\href{http://arxiv.org/abs/1403.5035}{{\ttfamily arXiv:1403.5035 [hep-th]}}.

\bibitem{Donnelly:2016auv}
W.~Donnelly and L.~Freidel, ``{Local subsystems in gauge theory and gravity},''
  \href{http://dx.doi.org/10.1007/JHEP09(2016)102}{{\em JHEP} {\bfseries 09}
  (2016) 102},
\href{http://arxiv.org/abs/1601.04744}{{\ttfamily arXiv:1601.04744 [hep-th]}}.

\bibitem{Gross:1992tu}
D.~J. Gross, ``{Two-dimensional QCD as a string theory},''
  \href{http://dx.doi.org/10.1016/0550-3213(93)90402-B}{{\em Nucl. Phys.}
  {\bfseries B400} (1993) 161--180},
\href{http://arxiv.org/abs/hep-th/9212149}{{\ttfamily arXiv:hep-th/9212149
  [hep-th]}}.

\bibitem{Gross:1993hu}
D.~J. Gross and W.~Taylor, ``{Two-dimensional QCD is a string theory},''
  \href{http://dx.doi.org/10.1016/0550-3213(93)90403-C}{{\em Nucl. Phys.}
  {\bfseries B400} (1993) 181--208},
\href{http://arxiv.org/abs/hep-th/9301068}{{\ttfamily arXiv:hep-th/9301068
  [hep-th]}}.

\bibitem{Ramgoolam:1993hh}
S.~Ramgoolam, ``{Comment on two-dimensional $O(N)$ and $Sp(N)$ Yang-Mills
  theories as string theories},''
  \href{http://dx.doi.org/10.1016/0550-3213(94)90237-2}{{\em Nucl. Phys.}
  {\bfseries B418} (1994) 30--44},
\href{http://arxiv.org/abs/hep-th/9307085}{{\ttfamily arXiv:hep-th/9307085
  [hep-th]}}.

\bibitem{Baez:1994gk}
J.~Baez and W.~Taylor, ``Strings and two-dimensional {QCD} for finite {$N$},''
  \href{http://dx.doi.org/10.1016/0550-3213(94)90125-2}{{\em Nucl. Phys.}
  {\bfseries B426} (1994) 53--70},
\href{http://arxiv.org/abs/hep-th/9401041}{{\ttfamily arXiv:hep-th/9401041
  [hep-th]}}.

\bibitem{Baez:1993gm}
J.~C. Baez, ``Strings, loops, knots and gauge fields,'' in {\em Knots and
  Quantum Gravity Riverside, California, May 14-16, 1993}.
\newblock 1993.
\newblock
\href{http://arxiv.org/abs/hep-th/9309067}{{\ttfamily arXiv:hep-th/9309067
  [hep-th]}}.
\newblock

\bibitem{Minahan:1993np}
J.~A. Minahan and A.~P. Polychronakos, ``{Equivalence of two-dimensional QCD
  and the $C = 1$ matrix model},''
  \href{http://dx.doi.org/10.1016/0370-2693(93)90504-B}{{\em Phys. Lett.}
  {\bfseries B312} (1993) 155--165},
\href{http://arxiv.org/abs/hep-th/9303153}{{\ttfamily arXiv:hep-th/9303153
  [hep-th]}}.

\bibitem{Gross:1993yt}
D.~J. Gross and W.~Taylor, ``{Twists and Wilson loops in the string theory of
  two-dimensional QCD},''
  \href{http://dx.doi.org/10.1016/0550-3213(93)90042-N}{{\em Nucl. Phys.}
  {\bfseries B403} (1993) 395--452},
\href{http://arxiv.org/abs/hep-th/9303046}{{\ttfamily arXiv:hep-th/9303046
  [hep-th]}}.

\bibitem{Jacobson:1994fp}
T.~Jacobson, ``{A Note on Hartle-Hawking vacua},''
  \href{http://dx.doi.org/10.1103/PhysRevD.50.R6031}{{\em Phys. Rev.}
  {\bfseries D50} (1994) R6031--R6032},
\href{http://arxiv.org/abs/gr-qc/9407022}{{\ttfamily arXiv:gr-qc/9407022
  [gr-qc]}}.

\bibitem{Cordes:1994sd}
S.~Cordes, G.~W. Moore, and S.~Ramgoolam, ``{Large N 2D Yang-Mills theory and
  topological string theory},''
  \href{http://dx.doi.org/10.1007/s002200050102}{{\em Commun. Math. Phys.}
  {\bfseries 185} (1997) 543--619},
\href{http://arxiv.org/abs/hep-th/9402107}{{\ttfamily arXiv:hep-th/9402107
  [hep-th]}}.

\bibitem{Taylor:1994zm}
W.~Taylor, ``{Counting strings and phase transitions in 2D QCD},''
\href{http://arxiv.org/abs/hep-th/9404175}{{\ttfamily arXiv:hep-th/9404175
  [hep-th]}}.

\bibitem{Polchinski:1994fq}
J.~Polchinski, ``{Combinatorics of boundaries in string theory},''
  \href{http://dx.doi.org/10.1103/PhysRevD.50.R6041}{{\em Phys. Rev.}
  {\bfseries D50} (1994) 6041--6045},
\href{http://arxiv.org/abs/hep-th/9407031}{{\ttfamily arXiv:hep-th/9407031
  [hep-th]}}.

\bibitem{Douglas:1993iia}
M.~R. Douglas and V.~A. Kazakov, ``{Large N phase transition in continuum QCD
  in two-dimensions},''
  \href{http://dx.doi.org/10.1016/0370-2693(93)90806-S}{{\em Phys. Lett.}
  {\bfseries B319} (1993) 219--230},
\href{http://arxiv.org/abs/hep-th/9305047}{{\ttfamily arXiv:hep-th/9305047
  [hep-th]}}.

\bibitem{Horava:1993aq}
P.~Ho\v{r}ava, ``{Topological strings and QCD in two dimensions},'' in {\em
  NATO Advanced Research Workshop on New Developments in String Theory,
  Conformal Models and Topological Field Theory Cargese, France, May 12-21,
  1993}.
\newblock 1993.
\newblock
\href{http://arxiv.org/abs/hep-th/9311156}{{\ttfamily arXiv:hep-th/9311156
  [hep-th]}}.
\newblock

\bibitem{Lee:2010ub}
S.-S. Lee, ``{Holographic description of large $N$ gauge theory},''
  \href{http://dx.doi.org/10.1016/j.nuclphysb.2011.05.011}{{\em Nucl. Phys.}
  {\bfseries B851} (2011) 143--160},
\href{http://arxiv.org/abs/1011.1474}{{\ttfamily arXiv:1011.1474 [hep-th]}}.

\bibitem{Dong:2016eik}
X.~Dong, D.~Harlow, and A.~C. Wall, ``{Reconstruction of Bulk Operators within
  the Entanglement Wedge in Gauge-Gravity Duality},''
  \href{http://dx.doi.org/10.1103/PhysRevLett.117.021601}{{\em Phys. Rev.
  Lett.} {\bfseries 117} no.~2, (2016) 021601},
\href{http://arxiv.org/abs/1601.05416}{{\ttfamily arXiv:1601.05416 [hep-th]}}.

\bibitem{Faulkner:2013ana}
T.~Faulkner, A.~Lewkowycz, and J.~Maldacena, ``{Quantum corrections to
  holographic entanglement entropy},''
  \href{http://dx.doi.org/10.1007/JHEP11(2013)074}{{\em JHEP} {\bfseries 11}
  (2013) 074},
\href{http://arxiv.org/abs/1307.2892}{{\ttfamily arXiv:1307.2892 [hep-th]}}.

\bibitem{Harlow:2016vwg}
D.~Harlow, ``{The Ryu-Takayanagi Formula from Quantum Error Correction},''
\href{http://arxiv.org/abs/1607.03901}{{\ttfamily arXiv:1607.03901 [hep-th]}}.

\end{thebibliography}\endgroup

\end{document}